\documentclass[twocolumn]{aastex63}
\newcommand{\squiggle}{SQuIGG$\vec{L}$E \,}

\usepackage{amsmath} 
\newcommand{\angstrom}{\text{\normalfont\AA}}

\usepackage{float}
\usepackage{textcomp}

\pdfoutput=1

\date{\today}

\submitjournal{ApJ}

\shorttitle{Flat Age Gradients in Massive z$\sim$0.6 Post-Starburst Galaxies}

\shortauthors{Setton et al.}

\begin{document}

\title{\squiggle Survey: Massive z$\sim$0.6 Post-Starburst Galaxies Exhibit Flat Age Gradients}

\author[0000-0003-4075-7393]{David J. Setton}
\affiliation{Department of Physics and Astronomy and PITT PACC, University of Pittsburgh, Pittsburgh, PA 15260, USA}

\author[0000-0001-5063-8254]{Rachel Bezanson}
\affiliation{Department of Physics and Astronomy and PITT PACC, University of Pittsburgh, Pittsburgh, PA 15260, USA}

\author[0000-0002-1714-1905]{Katherine A. Suess}
\affiliation{Astronomy Department, University of California, Berkeley, CA 94720, USA}

\author[0000-0002-4669-0209]{Qiana Hunt}
\affiliation{Department of Astronomy, University of Michigan, Ann Arbor, MI 48109, USA}

\author{Jenny E. Greene}
\affiliation{Department of Astrophysical Sciences, Princeton University, Princeton, NJ 08544, USA}

\author{Mariska Kriek}
\affiliation{Astronomy Department, University of California, Berkeley, CA 94720, USA}

\author[0000-0003-3256-5615]{Justin~S.~Spilker}
\altaffiliation{NHFP Hubble Fellow}
\affiliation{Department of Astronomy, University of Texas at Austin, 2515 Speedway, Stop C1400, Austin, TX 78712, USA}

\author[0000-0002-1109-1919]{Robert Feldmann}
\affiliation{Institute for Computational Science, University of Zurich, CH-8057 Zurich, Switzerland}

\author[0000-0002-7064-4309]{Desika Narayanan}
\affiliation{Department of Astronomy, University of Florida, 211 Bryant Space Sciences Center, Gainesville, FL 32611, USA}

\begin{abstract}

We present Gemini GMOS IFU observations of six massive ($M_\star\geq10^{11} \ M_\odot$) A-star dominated post-starburst galaxies at $z\sim0.6$. These galaxies are a subsample of the \squiggle Survey, which selects intermediate-redshift post-starbursts from the Sloan Digital Sky Survey spectroscopic sample (DR14) with spectral shapes that indicate they have recently shut off their primary epoch of star formation. Using $H\delta_A$ absorption as a proxy for stellar age, we constrain five of the galaxies to have young ($\sim 600$ Myr) light-weighted ages at all radii and find that the sample on average has flat age gradients. We examine the spatial distribution of mass-weighted properties by fitting our profiles with a toy model including a young, centrally concentrated burst superimposed on an older, extended population. We find that galaxies with flat $H\delta_A$ profiles are inconsistent with formation via a central secondary starburst. This implies that the mechanism responsible for shutting off this dominant episode of star formation must have done so uniformly throughout the galaxy.

\end{abstract}


\section{Introduction} \label{sec:intro}

Modern astronomical surveys have confirmed that the population of galaxies is bimodal, dividing fairly neatly into star-forming and quiescent populations. This bimodality is present in galaxy colors \citep[e.g.][]{Blanton2003, Jin2014}, sizes and structures \citep[e.g.][]{Shen2003,VanDerWel2014}, and star formation rates \citep[e.g.][]{Noeske2007}, although some studies suggest that the quiescent galaxies represent a long tail in star-formation rates as opposed to a distinct population and that the existence of a ``green valley" may be the result of optical sample selection \citep{Eales2018, Davies2019}. Nevertheless, at some point, all quiescent galaxies must have been star-forming; therefore, they are the descendants of a past star forming population that has turned off its star formation, or quenched. Whether the star forming progenitors of today's elliptical galaxies resembled their counterparts today remains to be seen \citep{Tadaki2020}. Despite this uncertainty, the existence of massive quiescent galaxies as early as $z\sim4$ \citep{Straatman2014, Davidzon2017,Tanaka2019, Forrest2020b, McLeod2020} implies that many galaxies transition from star-forming to quiescent via a rapid channel that shuts off star formation quickly and efficiently.

In order to empirically understand this rapid channel of quenching, one can study galaxies which have recently ended an intense episode of star formation. These galaxies, often called post-starburst (PSB) galaxies, can be identified by their spectral energy distributions (SEDs) which exhibit spectral shapes and features characteristic of A stars, indicating that star formation shut down in the last $\sim1$ Gyr \citep[e.g.][]{Dressler1983, Zabludoff1995}. Post-starburst galaxies can be selected by their strong Balmer absorption features and low star-formation rates \citep[e.g.][]{French2015}. While some post-starbursts appear to be galaxies in transition from star-forming to quiescent for the first time \citep{Alatalo2014}, others may be older ``K+A" quiescent galaxies with composite SEDs that include light from K-giants along with A stars formed in a frosting of recent star formation \citep{French2018a}.

Many physical models have been proposed for shutting off star formation and the relative efficiency of different physical mechanisms can vary throughout a galaxy. Therefore, the distribution of stellar ages in a galaxy, which themselves hold a record of the star formation history, can be used to distinguish among models. For example, mergers can drive gas to the center of a galaxy, triggering a strong burst of star-formation, after which the galaxy quenches \citep{Hopkins2008,Snyder2011, Wellons2015}. This quenching pathway would produce a positive age gradient, where the stellar population at the center of the galaxy is younger than the population on the outskirts. In contrast, simulations of quenching via wet compaction, where gas migrates inward in a way which compacts a galaxy, suggest that galaxies may experience extended star formation outside their core after the central gas is depleted, resulting in negative radial age gradients \citep{Tacchella2015a, Zolotov2015}.

Spatially resolved studies of the stellar populations of post-starburst galaxies at low-redshift have found a range of stellar age profiles traced by spectral indicators like $H\delta_A$ and $\mathrm{D_n}4000$ \citep{Yagi2006, Pracy2005, Chen2019}, and local starburst galaxies appear to be experiencing centrally concentrated bursts \citep{Ellison2020}. However, these galaxies may not be representative of the evolutionary path that quenches galaxies for the first time. While low-mass local post-starburst galaxies can be extremely burst-dominated, at higher mass where the aforementioned bimodalities are the most extreme, post-starbursts are predominantly ``K+A" post-starbursts in which a small burst has occurred in a quiescent galaxy that formed at high redshift \citep{Helmboldt2008, French2018a}. This indicates that while massive post-starbursts do exist in the local universe, they do not tend to be galaxies that are quenching their primary epoch of star formation, which is unsurprising given that local massive galaxies are almost exclusively old \citep[e.g.][]{McDermid2015}. Furthermore, post-starbursts constitute a negligible part of the $z<1$ luminous galaxy population \citep{Pattarakijwanich2016} and while massive post-starbursts exist at intermediate redshift ($0.5<z<1$), it is not until $z=2$ that they start to represent a significant fraction of the population of massive, quenched galaxies \citep{Whitaker2012a,Wild2016}. Thus, in order to understand the galaxies which are quenching their primary epoch of star formation, we must look to earlier cosmic time.

While $z=2$ quenched galaxies are still beyond the reach of spatially resolved spectroscopic studies outside of extreme lensed systems \citep[e.g.][]{Jafariyazani2020, Akhshik2020}, intermediate-redshift post-starbursts are more accessible. Post-starburst galaxies have been studied in the Large Extra Galactic Astrophysics Census (LEGA-C) survey, which consists of deep ($\sim$20 hours/galaxy) spectra of galaxies at $z\sim0.8$ in the COSMOS field \citep{VanDerWel2016}. These intermediate mass ($10^{10} \ M_\odot<M_\star<10^{11} \ M_\odot$) post-starbursts have positive age gradients \citep{DeugenioF2020} and compact sizes \citep{Wu2018, Wu2020} consistent with formation via a recent central starburst. However, these galaxies, like their counterparts at low redshift, are observed following a frosting of recent star formation; the strong, but not extreme, $H\delta_A$ in their sample indicates that K-giant stars are contributing significantly to the optical light of these galaxies. LEGA-C's pencil beam survey design does not allow it to find the rare but crucial A-star dominated post-starbursts that are in the stage of rapid transition from star forming to quiescent. By leveraging the wide-area of the Sloan Digital Sky Survey (SDSS), one can identify rare galaxies at intermediate redshift that have recently shut off their primary epoch of star formation. In a pilot program, \citet{Hunt2018} found a flat age gradient in a single z=0.747 galaxy, indicating that the most extreme post-starbursts may quench differently than their less extreme ``K+A" counterparts.

\begin{figure*}
\includegraphics[width=\textwidth]{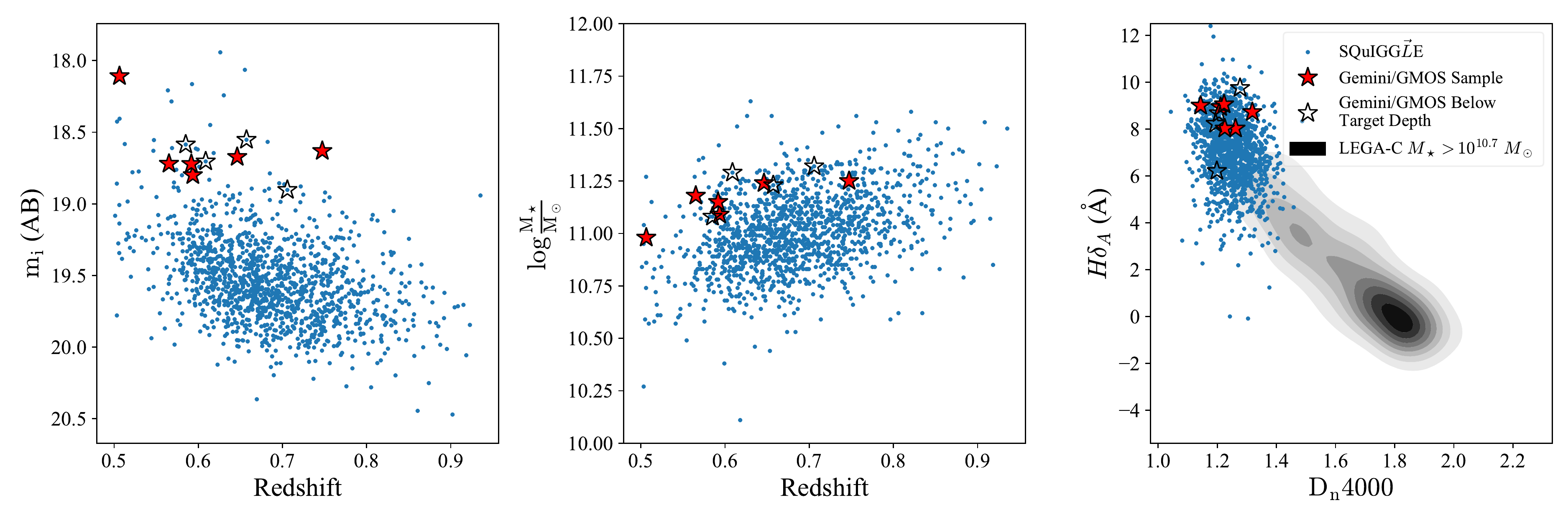}
\caption{SDSS observed i-band magnitude versus redshift (left), stellar mass versus redshift (center), and $H\delta_A$ versus $\mathrm{D_n 4000}$ (right) for the entire \squiggle sample. In each panel, the Gemini/GMOS targets are shown as stars, where filled stars are those that are included in this sample. The Gemini/GMOS targets are among the brightest galaxies in the parent sample by selection. In the right panel, the  distribution of similarly massive ($M_\star>10^{10.7} \ M_\odot$) galaxies at $0.6<z<0.8$ from the LEGA-C survey are indicated by the grey shaded region. The \squiggle galaxies are all significantly offset in $H\delta_A$ versus $\mathrm{D_n4000}$ from the LEGA-C quiescent galaxies at this redshift. The Gemini/GMOS targets span the \squiggle range of $\mathrm{D_n4000}$ and are higher on average in $H\delta_A$. \label{fig:selection}}
\end{figure*}

In this work, we build on that study of J0912+1523 and present five additional IFU observations of massive ($M_\star\geq10^{11} M_\odot$), burst-dominated post-starburst galaxies at z $\sim$ 0.6. In Section \ref{sec:data}, we describe the parent \squiggle sample, as well as our spectroscopic analysis of the follow-up GMOS observations presented in this work. In Section \ref{sec:analysis}, we discuss the spatially resolved stellar populations of the sample. Finally, in Section \ref{sec:discussion}, we highlight the implications of our study on the quenching of massive galaxies in this epoch. Throughout this paper we assume a concordance $\Lambda$CDM cosmology with $\Omega_{\Lambda}=0.7$, $\Omega_m=0.3$ and $H_0=70$ $\mathrm{km\,s^{-1}\,Mpc^{-1}}$, and quote AB magnitudes.

\begin{deluxetable*}{cccccccccc}
\tabletypesize{\scriptsize}
\tablecaption{Properties of the GMOS observations\label{tbl:gmos}}

\tablehead{
\colhead{ID} & \colhead{Name} & \colhead{RA} & \colhead{Dec} & \colhead{z} & \colhead{Stellar} & \colhead{i} &  \colhead{GMOS} & \colhead{Integration\tablenotemark{1}} & \colhead{Image\tablenotemark{2}}\\[-0.4cm]
\colhead{} & \colhead{} & \colhead{} & \colhead{} & \colhead{} & \colhead{Mass} & \colhead{} & \colhead{Program} & \colhead{Time} & \colhead{Quality} \\[-0.2cm]
\colhead{} & \colhead{} & \colhead{(degrees)} & \colhead{(degrees)} & \colhead{} & \colhead{($\mathrm{log\frac{M_{\star}}{M_{\odot}}}$)} & \colhead{(AB Mag)} &  \colhead{} & \colhead{(s)} & \colhead{} 
}

\startdata
SDSS J110932.14-004003.8 & J1109-0040 & 167.384 & -0.6678 & 0.593 & 11.09& 18.8 & GN-2019A-Q-234 & 9720 & 20 \\
SDSS J023359.33+005238.4 & J0233+0052 & 38.4972 & 0.8774 & 0.592 & 11.15& 18.72 & GN-2017B-Q-37 & 7560 (6480) & 20 \\
SDSS J091242.76+152305.1 & J0912+1523 & 138.1782 & 15.3848 & 0.747 & 11.25& 18.63 & GN-2016A-FT-6 & 9802 & 20 \\
SDSS J083547.08+312144.5 & J0835+3121 & 128.9462 & 31.3624 & 0.506 & 10.98& 18.11 & GN-2017B-Q-37 & 6300 & 70 \\
SDSS J075344.17+240336.1 & J0753+2403 & 118.4341 & 24.0601 & 0.565 & 11.18& 18.72 & GN-2017B-Q-37 & 8640 (5400) & 70 \\
SDSS J144845.91+101010.5 & J1448+1010 & 222.1913 & 10.1696 & 0.646 & 11.24& 18.67 & GS-2018A-FT-112 & 11880 (8640) & 70 
\enddata

\tablenotetext{1}{Integration time that was considered useful are shown in parentheses if they differ from the total integration time. Useful frames of data are those which do not have visible issues after scattered light subtraction of significant noise spikes in the $H\delta_A$ bandpass.}
\tablenotetext{2}{The Image Quality quoted is the worst case conditions under which the majority of frames were observed. IQ20 corresponds to FWHM$\leq$0.5" seeing and IQ70 corresponds to FWHM$\leq$0.75" seeing at zenith.}
\end{deluxetable*}

\begin{figure*}
\includegraphics[width=\textwidth]{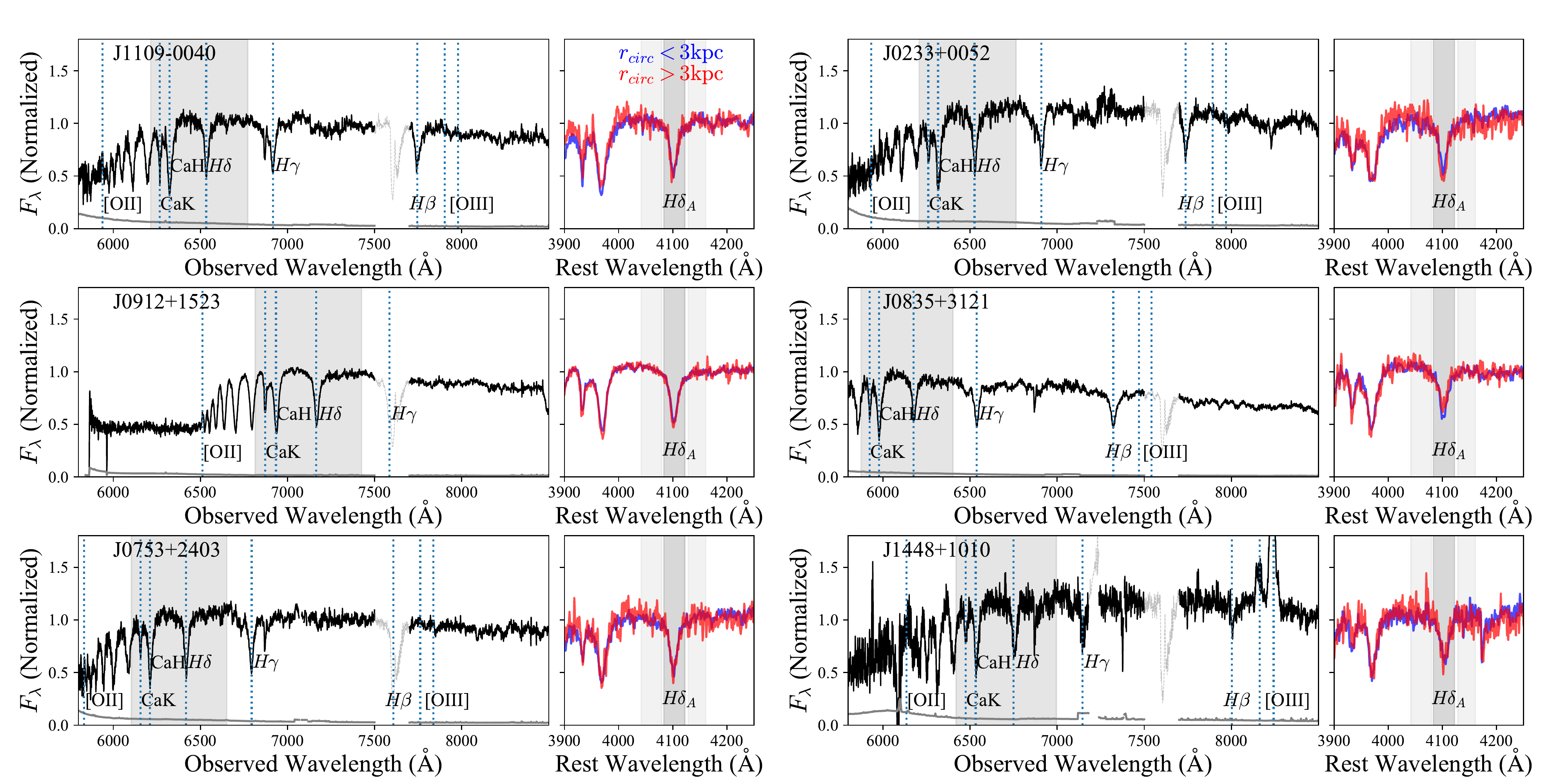}
\caption{The Gemini/GMOS spectra integrated within a 2" circular aperture for each target are shown on the left. The grey line is the error on the flux and relevant lines are labeled. The spectra are plotted as transparent where they are masked due to strong telluric features. The luminosity-weighted inner ($r_{circ}<$3 kpc, blue) and outer ($r_{circ}>$3 kpc, red) spectra are shown in the right panel, highlighting the Lick $H\delta_A$ bandpass (dark gray) and continuum (light gray). Absorption features are remarkably uniform in the inner and outer spectra, with strong Balmer absorption throughout. All spectra are normalized to the flux at 4000 $\angstrom$ in the rest-frame. \label{fig:images_and_spectra}}
\end{figure*}

\section{Data} \label{sec:data}

\subsection{The \squiggle Sample}

For this study, we target a subsample of galaxies from the \squiggle (Studying Quenching in Intermediate-z Galaxies: Gas, Angu$\vec{L}$ar Momentum, and Evolution) Survey (K. Suess et. al in preparation). \squiggle is designed to identify post-starburst galaxies that have recently quenched their primary epoch of star formation. The survey identifies all galaxies at $z>0.5$ with integrated signal-to-noise of 6 in synthetic rest-frame U, B, and V filters from spectroscopic data in SDSS DR14 \citep{Abolfathi2018}. We use the rest-frame color cuts ($U-B>0.975; -0.25<B-V<0.45$) in \citet{Kriek2010} to identify post-starburst galaxies with strong Balmer breaks and blue colors redward of the break, thereby selecting A-star dominated spectral energy distributions (SEDs). This selection identifies 1318 unique galaxies with $0.5<z<0.94$ and $17.94<i<20.47$. To characterize the stellar populations and measure stellar masses, we perform stellar population synthesis modeling of the SDSS spectra and \textit{ugriz} photometry using \texttt{FAST++}\footnote{\href{https://github.com/cschreib/fastpp}{https://github.com/cschreib/fastpp}}, an implementation of the popular \texttt{FAST} program \citep{Kriek2009}. We assume a delayed exponential star formation history, BC03 stellar population libraries \citep{Bruzual2003}, a \citet{Chabrier2003} initial mass function, and a \citet{Calzetti1997} dust law. The galaxy masses span $10^{10.11} M_\odot<M_\star<10^{11.63} M_\odot$ with mean $M_\star=10^{11.0} M_\odot$. Although \squiggle galaxies are not explicitly selected based on their $H\delta_A$ absorption, the sample exclusively exhibits strong Balmer absorption consistent with the common post-starburst selection: 98\% of the sample meets a $H\delta_A>4 \ \angstrom$ criterion sometimes used to select post-starburst galaxies \citep[e.g.][]{French2015, Wu2018}. The extremely strong Balmer absorption in this sample (median $H\delta_A$=7.12 $\angstrom$) reflects a recently burst of star formation is dominating both the mass and light of these galaxies \citep[e.g.][]{Kauffmann2003a}. 

\subsection{Gemini/GMOS Observations} \label{subsec:reduction}

From the 1318 \squiggle galaxies, we conducted follow-up observations of ten optically-bright \squiggle galaxies using the GMOS IFU instruments on Gemini North and South. In Figure \ref{fig:selection}, we show the Gemini/GMOS targets (large stars) and the parent \squiggle sample (small symbols) in SDSS i magnitude versus redshift, mass versus redshift, and $H\delta_A$ versus $\mathrm{D_n4000}$. Comparing to massive galaxies in the LEGA-C survey, the high $H\delta_A$ and low $\mathrm{D_n4000}$ of \squiggle indicate that the post-starburst galaxies we select are indeed significantly younger than typical $z\sim0.7$ quiescent galaxies. Objects that fall short of target depths are indicated by open symbols. The integration times ($\sim2.5$ hours/galaxy) were chosen to measure the stellar continuum in spatially resolved spaxels and annuli to probe the kinematics and ages of the stellar populations of each galaxy. Each galaxy was observed using the R400 grating ($5500 \angstrom<\lambda<10500 \angstrom$). The observations were collected between 2016 and 2019. Each exposure was bias subtracted, scattered light corrected, cosmic ray rejected, flat field corrected, wavelength calibrated, response corrected (using a standard star which was not observed on the same night as the observations), and sky subtracted using the gfreduce and related IRAF packages following \citet{Lena2014}. All individual datacubes were constructed using the gfcube package at a resolution of 0.05 "/pixel. Each spatial pixel was then iteratively sigma clipped using \texttt{astropy} sigma\_clip to remove noise spikes at the 5-$\sigma$ level \citep{astropy2013, astropy2018}. We performed an additional sky subtraction using a spline fit to the median of the outer pixels to account for any catastrophic over- or under-subtraction of the continuum in the GMOS pipeline.

\begin{figure*}
\includegraphics[width=\textwidth]{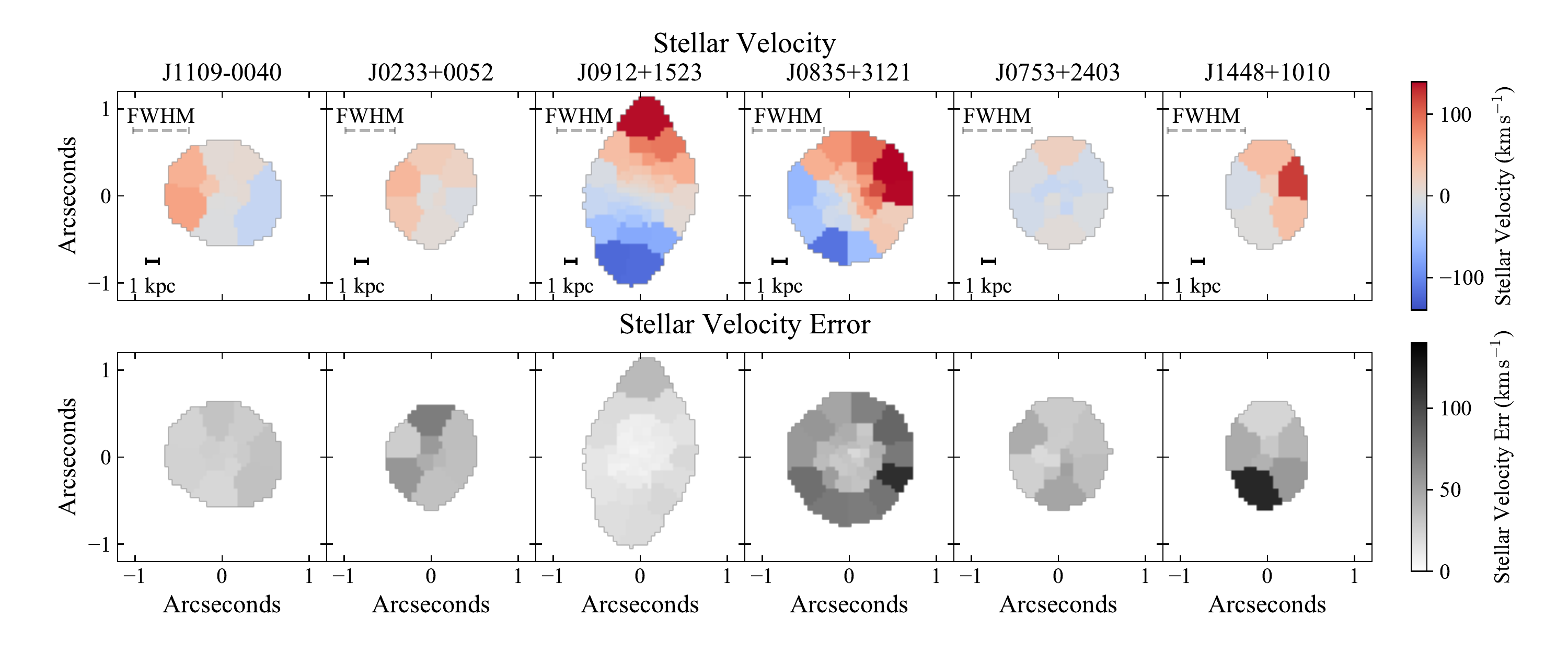}
\caption{Measured stellar velocity (top) and velocity error (bottom) maps for the massive post-starburst galaxies in this sample. In J1109+0040, J0912+1523, and J0835+3121, unambiguous velocity gradients are detected, despite very different seeing conditions. For the rest of the sample, ordered motion is either ambiguous or non-existent. Effective point spread functions are not well-constrained for these observations, but worst-case FWHM limits are indicated by gray dashed lines.
\label{fig:velocity_2d}}
\end{figure*}

We combined all individual reduced datacubes for each object using an inverse variance weighted average. We truncate the spectra above 8500 $\angstrom$ because the data quality drops off severely and there are no useful spectral features present. Because the standard stars were not observed under the same viewing conditions and orientation as the science frames, the initial response correction is uncertain. We use the SDSS spectra to improve the response correction, using the luminosity weighted average of each cube within the 2" or 3" SDSS fiber aperture to obtain a 1D integrated galaxy spectrum. We fit a 10th order polynomial to models of the SDSS continuum generated by \texttt{FAST}, and do the same with the integrated GMOS spectra. We use the ratio of these polynomials to rectify the spectral shape of each individual spaxel.

Individual spaxels in the datacubes generally have insufficient signal-to-noise to facilitate robust measurement of stellar absorption features (median S/N$\sim$0.8 $\angstrom^{-1}$ in the outer spaxels), so some binning is required. We Voronoi bin \citep{Cappellari2003, Cappellari2009} the data to a signal-to-noise of 6 $\angstrom^{-1}$ at $\sim4100 \  \angstrom$. In addition, we bin the spaxels using elliptical isophotes with an axis ratio and orientation we fit with the \texttt{photutils} python package \citep{photutils}. We adopt an adaptive binning scheme, expanding the semi-major axis of the isophotal ellipses until we reach a target uncertainty in the Lick $H\delta_A$ index less than 1.5 $\angstrom$. These annular measurements sacrifice spatial resolution to gain signal to noise and provide a natural comparison to radial models by use of the circularized radius, defined as $r_{circ}\equiv\sqrt{ab}$, where a and b are the semi-major and semi-minor axes of the ellipse that intersects the center of the bin

Unfortunately, the signal-to-noise in the stellar continuum in several datacubes were insufficient for this analysis. Three were not observed to a depth where we could resolve six Voronoi bins, so we exclude them from all additional analysis. We also exclude one other galaxy due to issues with strong residual sky features that overlap with $H\delta_A$ spectral feature. The details of the remaining six galaxies, including integration times and approximate seeing conditions, are presented in Table \ref{tbl:gmos}. The integrated spectra of the sample are shown in the left panels of Figure \ref{fig:images_and_spectra}. In the right panels, we highlight the $H\delta_A$ bandpass and show luminosity weighted spectra from inner ($r_{circ}<3 \ \mathrm{kpc}$, blue) and outer ($r_{circ}>3 \ \mathrm{kpc}$, red) annuli. In all galaxies except for J0835+3121, the Balmer absorption is similarly strong in both the inner and outer bins.

We use Penalized Pixel Fitting (\texttt{pPXF}) \citep{Cappellari2004} to measure the stellar line-of-sight velocities in each Voronoi bin. We fit the spectra using theoretical stellar spectral libraries to match the spectral resolution of the observations, which are $\sim0.5 \ \angstrom \ \mathrm{pixel}^{-1}$ in the rest frame \citep{Bezanson2018}. We fit the spatially binned spectra using a 1st order multiplicative polynomial and 5th order additive polynomial to account for uncertainty in the continuum shape. We measure the Lick $H\delta_A$ index using \texttt{pyphot}\footnote{\href{https://github.com/mfouesneau/pyphot}{https://github.com/mfouesneau/pyphot}}, fitting the continuum with a 1st order polynomial. We estimate the uncertainty in this index via a 1000-iteration Monte Carlo resampling of the error vector.

Due to the limited field of view of GMOS (3.5"x5"), our observations never include nearby stars, and as such, cannot exactly constrain the effective point spread function (PSF). However, due to the small angular sizes of the sample galaxies, we expect the impact of beam smearing to be strong, and must account for it in our analysis. Gemini provides seeing information in their RAW-IQ scores for the observations that measure an upper limit for the FWHM of the PSF at zenith. For galaxies observed in IQ-20, the seeing at zenith should be no greater than 0.5". For galaxies observed in IQ-70, the seeing at zenith should be no greater than 0.75". These values can be corrected for airmass effects as:
 
 \begin{equation}
\mathrm{FWHM}_{corr} = \mathrm{FWHM}_{zenith} * (\mathrm{airmass})^{0.6}
\end{equation}

\noindent We treat $\mathrm{FWHM_{corr}}$ as the FWHM of a Moffat profile and consider it to be a conservative upper limit on the seeing that we use in our analysis of these galaxies. 

\begin{figure*}
\plotone{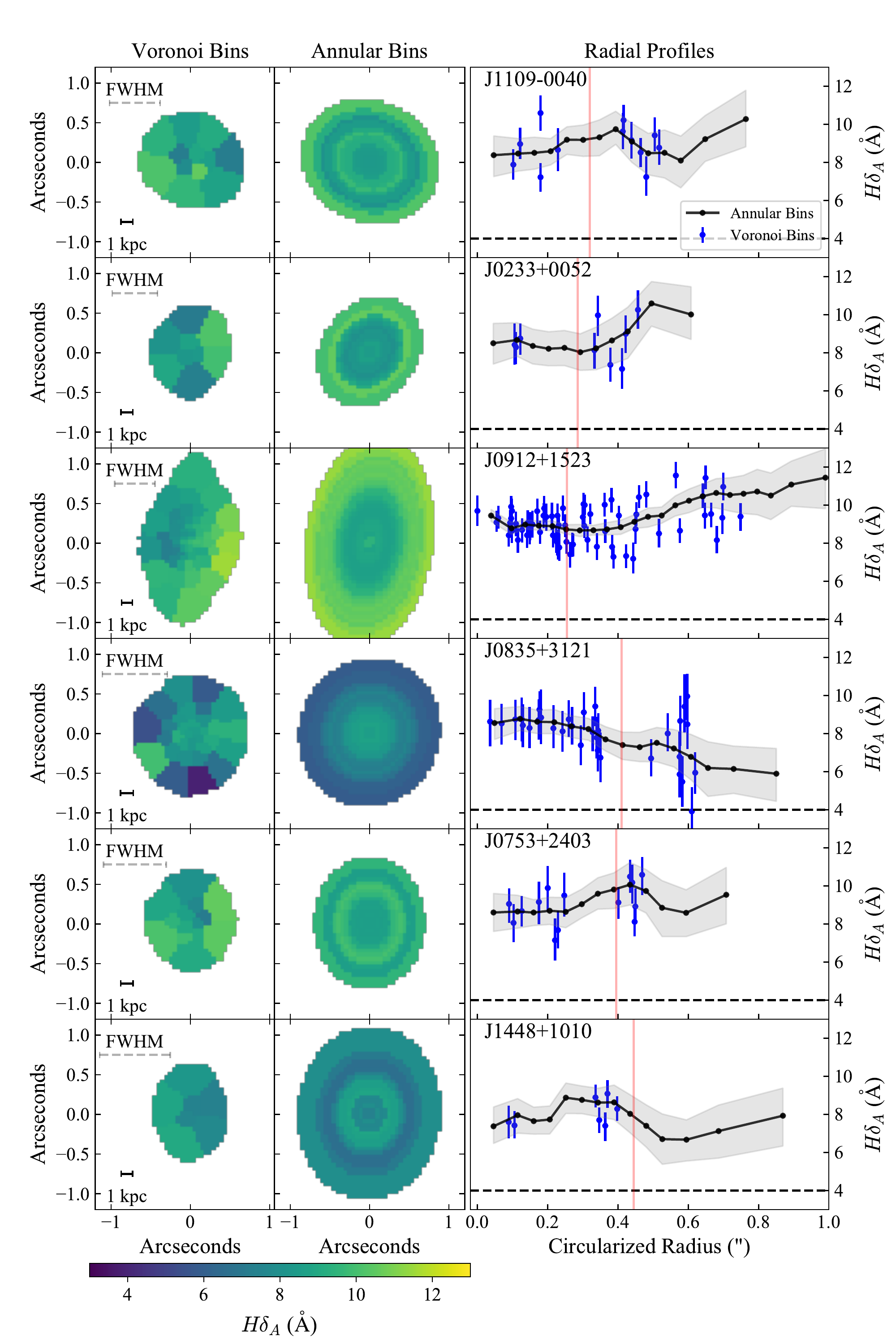}
\caption{$H\delta_A$ maps for the final sample in Voronoi (left) and annular bins (center). The dashed error bars represents the FWHM of the worst case seeing conditions. (Right): $H\delta_A$ profiles versus circularized radius for Voronoi bins (blue symbols) and annular bins (black line with gray band). The red vertical lines represent the worsts-case half-width half maximum of the point spread function. The dashed horizontal line at 4 $\angstrom$ indicates the common threshold to spectroscopically identify post-starburst galaxies. The entire sample is consistent with PSB-like absorption at all radii.
\label{fig:hdelta_2d}}
\end{figure*}

\begin{figure}
\plotone{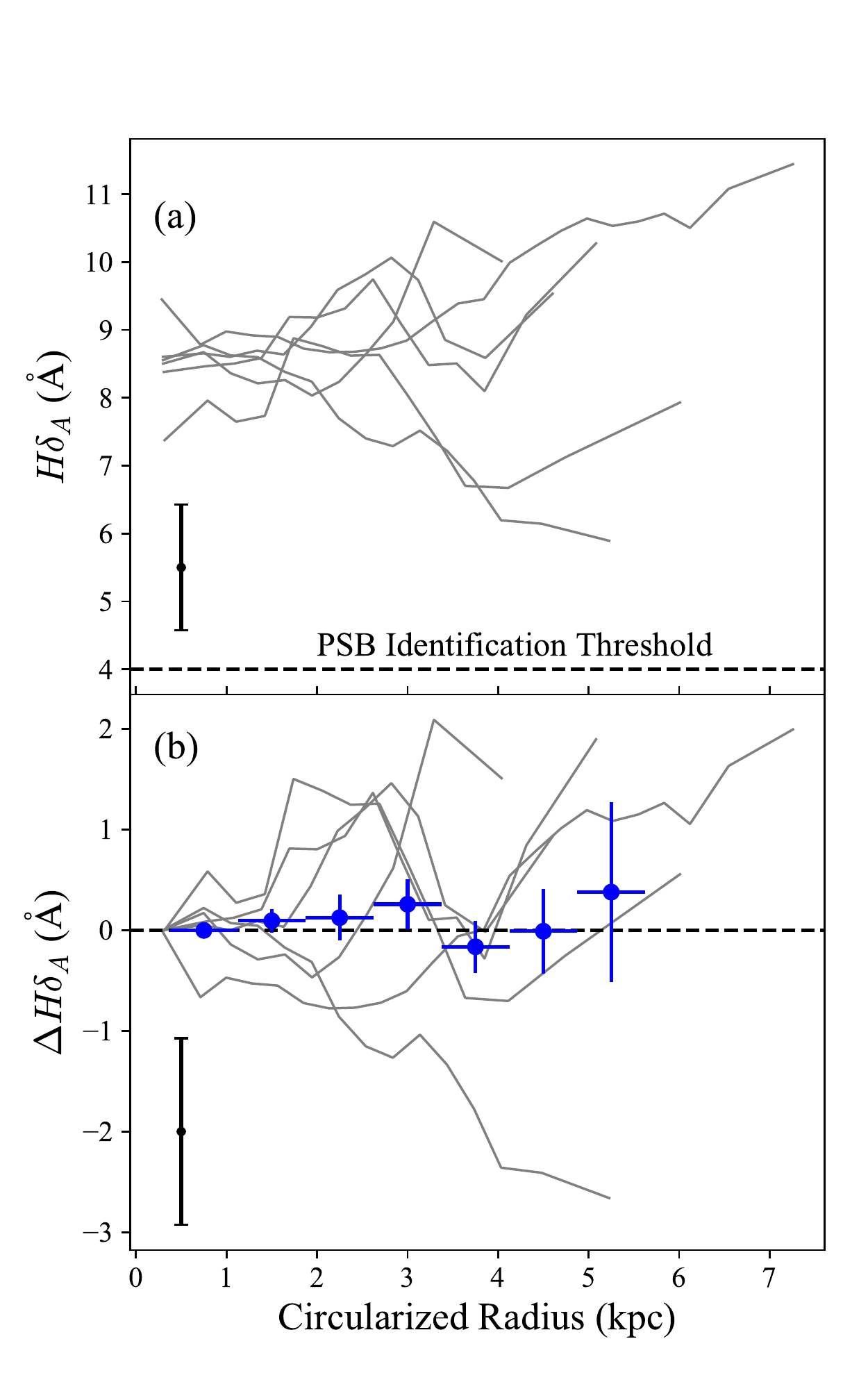}
\caption{(a) Annular $H\delta_A$ profiles (a) and gradients (b) for the sample as a function of physical radius. The error bar represents the average error in the measurements of $H\delta_A$. (b) Radial $H\delta_A$ trends relative to the central measurement. The blue points show the running mean and error on the mean in 0.75 kpc bins. The average profile is flat to 5.5 kpc. \label{fig:hdelta_all_gal}}
\end{figure}

\begin{figure*}
\includegraphics[width=\textwidth]{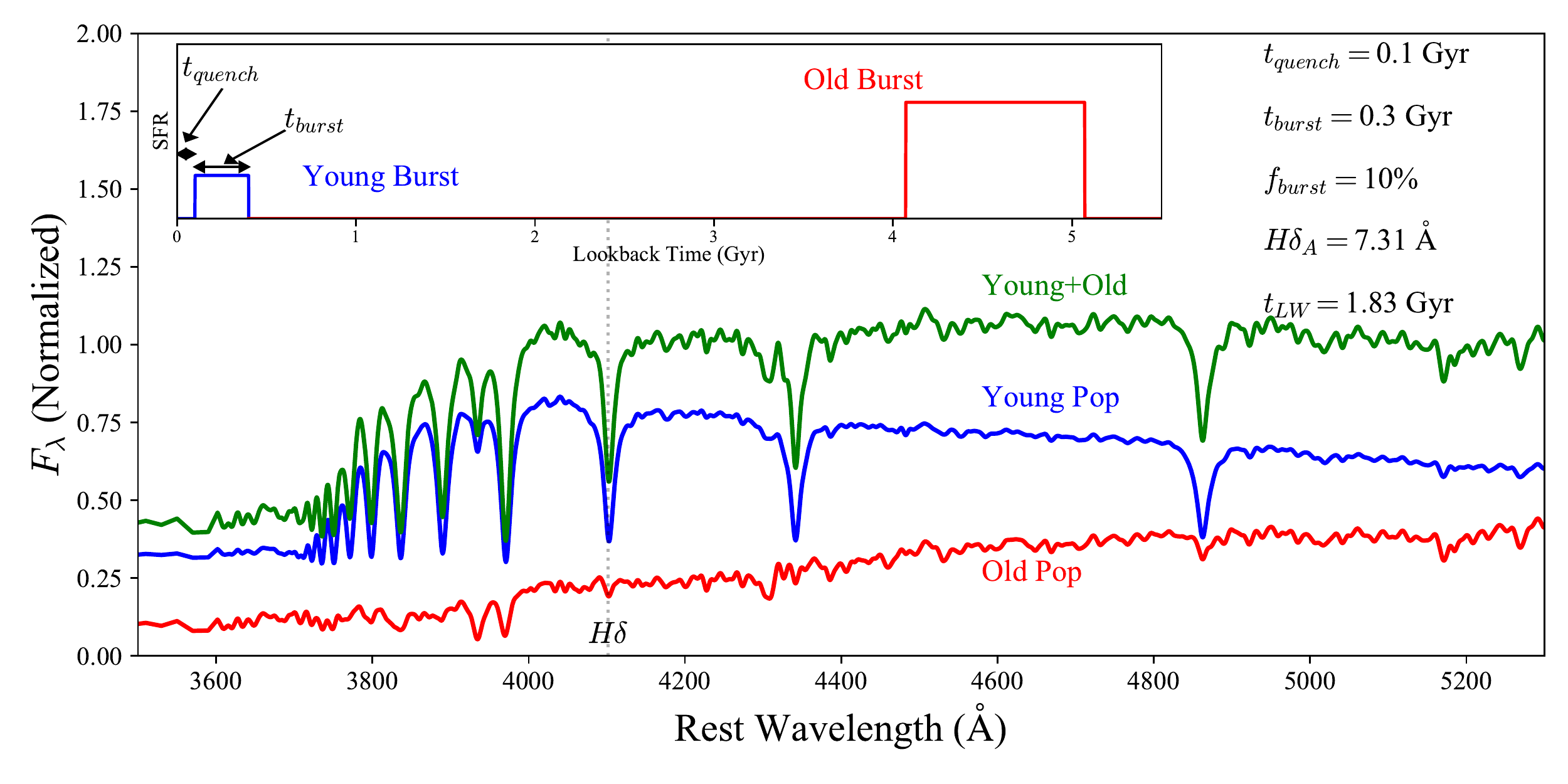}
\caption{An example of the model spectra that can be generated using the two burst toy model. In the main figure, old population that formed 90\% of the total mass at z=2, representing a population of older stars which formed at early times (red spectrum), is combined with a recent 300 Myr burst that quenched 100 Myr prior to observation (blue spectrum), to produce a post-starburst integrated spectrum (green). The spectra are normalized to the flux at 4000 $\angstrom$ in the composite spectrum. The inset panel shows the star formation history that produces these SEDs, with the same color scheme as the main figure. 
\label{fig:toy_model}}
\end{figure*}

\section{Analysis} \label{sec:analysis}

\subsection{Stellar Velocities and $H\delta_A$ Profiles}

Using the Voronoi bins and elliptical annuli discussed in Section \ref{subsec:reduction}, we can study the ordered motion and spatially-resolved age sensitive features in this sample of massive post-starbursts. In Figure \ref{fig:velocity_2d}, we present stellar velocity maps of the sample. Two galaxies (J0912+1523 and J0835+3121) show clear signs of strong ordered motion, with a third (J1109-0040) showing weaker but still significant velocity gradient. The other three galaxies do not exhibit statistically significant velocity gradients. Detailed analysis of the intrinsic velocity structures of these galaxies would require a more precise model of the point spread function and is outside the scope of the current paper (see \citet{Hunt2018} for more in depth discussion about the velocity structure of J0912+1523). However, the fact that we successfully resolve rotation in one of the galaxies that was observed under the worst seeing conditions indicates that all targets are at least marginally spatially resolved.

The $H\delta$ absorption feature at $4100 \ \angstrom$ and the $\mathrm{D_n4000}$ ratio of flux redward and blueward of the Balmer/$4000 \ \angstrom$ break together are very powerful in constraining the age of a stellar population \citep{Kauffmann2003a}. However, $\mathrm{D_n4000}$ is very sensitive to systematic uncertainties in the sky subtraction and response correction, especially in faint outer spaxels (for more, see \citet{Hunt2018}). In contrast, $H\delta_A$ is insensitive to both these uncertainties in addition to dust extinction. Given the extreme $H\delta_A$ exhibited by galaxies in this sample, we elect to use it alone as a tracer of stellar age, as strong absorption is still very constraining (see Section \ref{subsec:age_gradients}). In Figure \ref{fig:hdelta_2d}, we show $H\delta_A$ maps in both Voronoi (left) and annular (center) binning schemes. The right column shows sets of $H\delta_A$ measurements versus the circularized radius. In blue, the Voronoi measurements are plotted with associated errors. The black line and shaded region correspond to the annular measurements and associated errors. All galaxies exhibit PSB-like light ($H\delta_A>4 \ \angstrom$, black dashed line) at all radii. In Figure \ref{fig:hdelta_all_gal}, all annular profiles (\ref{fig:hdelta_all_gal}a) and gradients (\ref{fig:hdelta_all_gal}b) are shown as a function of the physical circularized radius with a characteristic error bar in the bottom corner. The average $\Delta H\delta_A$ is shown as blue points with error bars representing the error in the mean. Out to 5.5 kpc, the average gradient of this sample is flat. If we remove J0835+3121, which is host to the most significant gradient, there is a small signature of stronger absorption at large radii.

In Table \ref{tbl:hdelta_r}, we list the measurements of $H\delta_A$ and $H\delta_A$ gradients. The $H\delta_A$ indices measured from spatially integrated spectra in a 2" aperture are listed first with associated errors. We also measure the slope of the annular profiles using the publicly available Markov Chain Monte Carlo (MCMC) fitting code emcee \citep{Foreman-Mackey2013} to perform a linear regression. Four of the six galaxies are consistent with a flat $H\delta_A$ gradient at the 2-$\sigma$ level, while the galaxy J0912+1523 has a slightly increasing $H\delta_A$ profile. Only J0835+3121 exhibits a negative gradient in $H\delta_A$. We also measure the Spearman correlation coefficients for each $H\delta_A$ profile and reach similar conclusions to those in the linear regressions.

\begin{deluxetable}{cccc}
\tabletypesize{\scriptsize}
\tablecaption{Properties of Radial $H\delta_A$\label{tbl:hdelta_r}}
\tablehead{
\colhead{Name} & \colhead{Integrated $H\delta_A$ \tablenotemark{1}} & \colhead{$\frac{dH\delta_A}{dr}$} & \colhead{Spearman $\rho$\tablenotemark{2}} \\[-0.2cm]
 \colhead{} & \colhead{($\angstrom)$} & \colhead{($\angstrom\,\mathrm{kpc}^{-1}$)} & 
 \colhead{}
 }

\startdata
J1109-0040 & 8.94 $\pm$ 0.15 & 0.21 $\pm$ 0.21 & 0.371 \\
J0233+0052 & 8.73 $\pm$ 0.17 & 0.47 $\pm$ 0.32 & 0.482 \\
J0912+1523 & 9.0 $\pm$ 0.04 & 0.25 $\pm$ 0.06 & 0.834 \\
J0835+3121 & 8.03 $\pm$ 0.13 & -0.66 $\pm$ 0.17 & -0.968 \\
J0753+2403 & 9.06 $\pm$ 0.15 & 0.3 $\pm$ 0.24 & 0.401 \\
J1448+1010 & 8.02 $\pm$ 0.25 & -0.03 $\pm$ 0.22 & -0.191
\enddata

\tablenotetext{1}{This value is measured on the luminosity weighted combination of spaxels within a 2"-diameter circular aperture.}
\tablenotetext{2}{The Spearman correlation coefficient for the annular measurements of $H\delta_A$ as a function of the circularized radius.}
\end{deluxetable}

\begin{figure*}
\includegraphics[width=\textwidth]{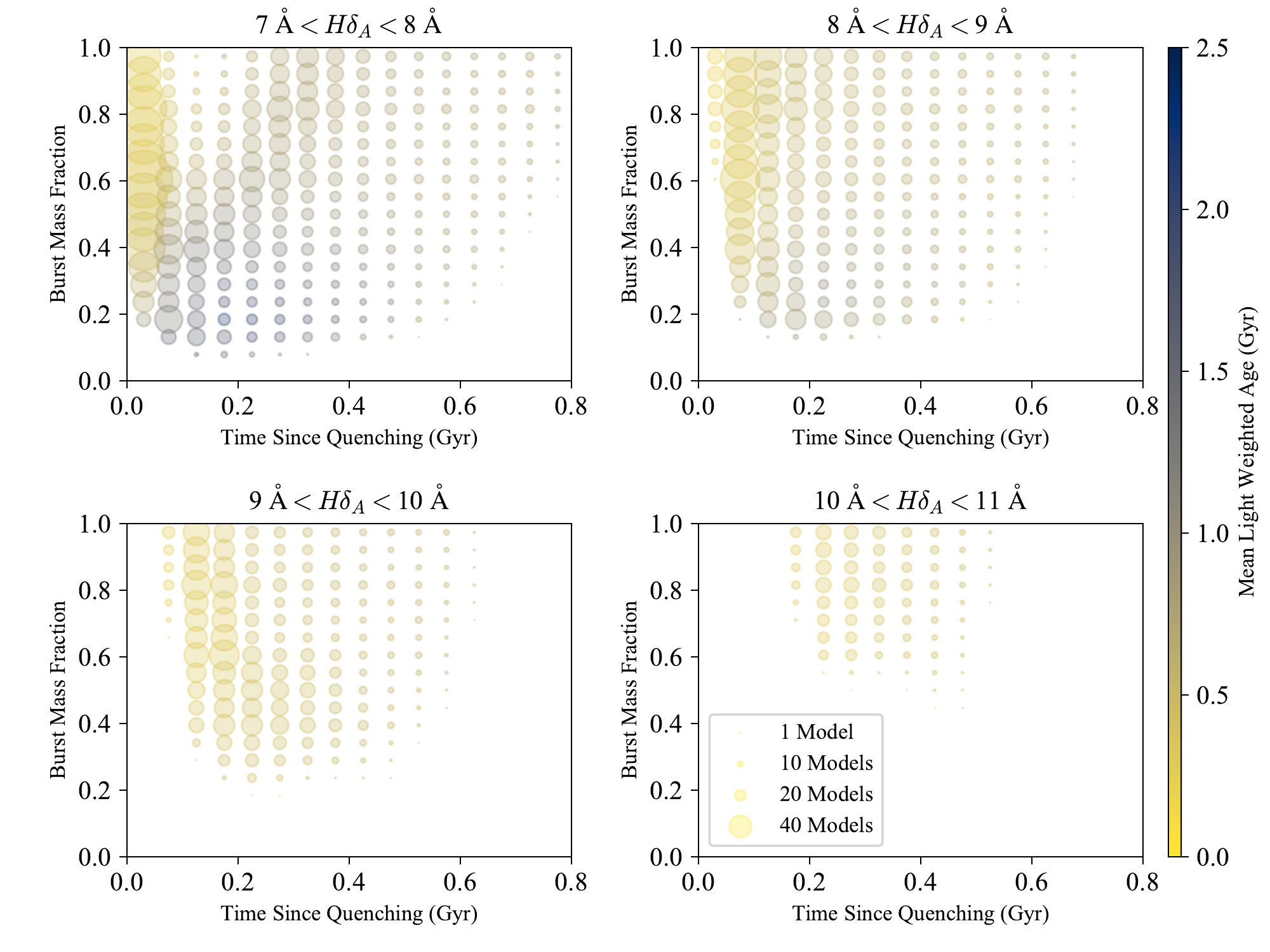}
\caption{The secondary burst fraction versus time since quenching for our toy models which result in $H\delta_A>7 \ \angstrom$. The points are colored by the light-weighted age (at 4000 $\angstrom$) and the size of the points is proportional to the number of models in the library that meet the $H\delta_A$ criteria in that bin. The plots are divided into the models which have $H\delta_A$ 7-8 $\angstrom$ (top left), 8-9 $\angstrom$ (top right), 9-10 $\angstrom$ (bottom left), and 10-11 $\angstrom$ (bottom right). High values of $H\delta_A$ are only possible for galaxies with light-weighted ages that are on the order of the time since quenching and an A-type stellar population is dominating the light.
\label{fig:binned_models}}
\end{figure*}

\begin{figure}[]
\plotone{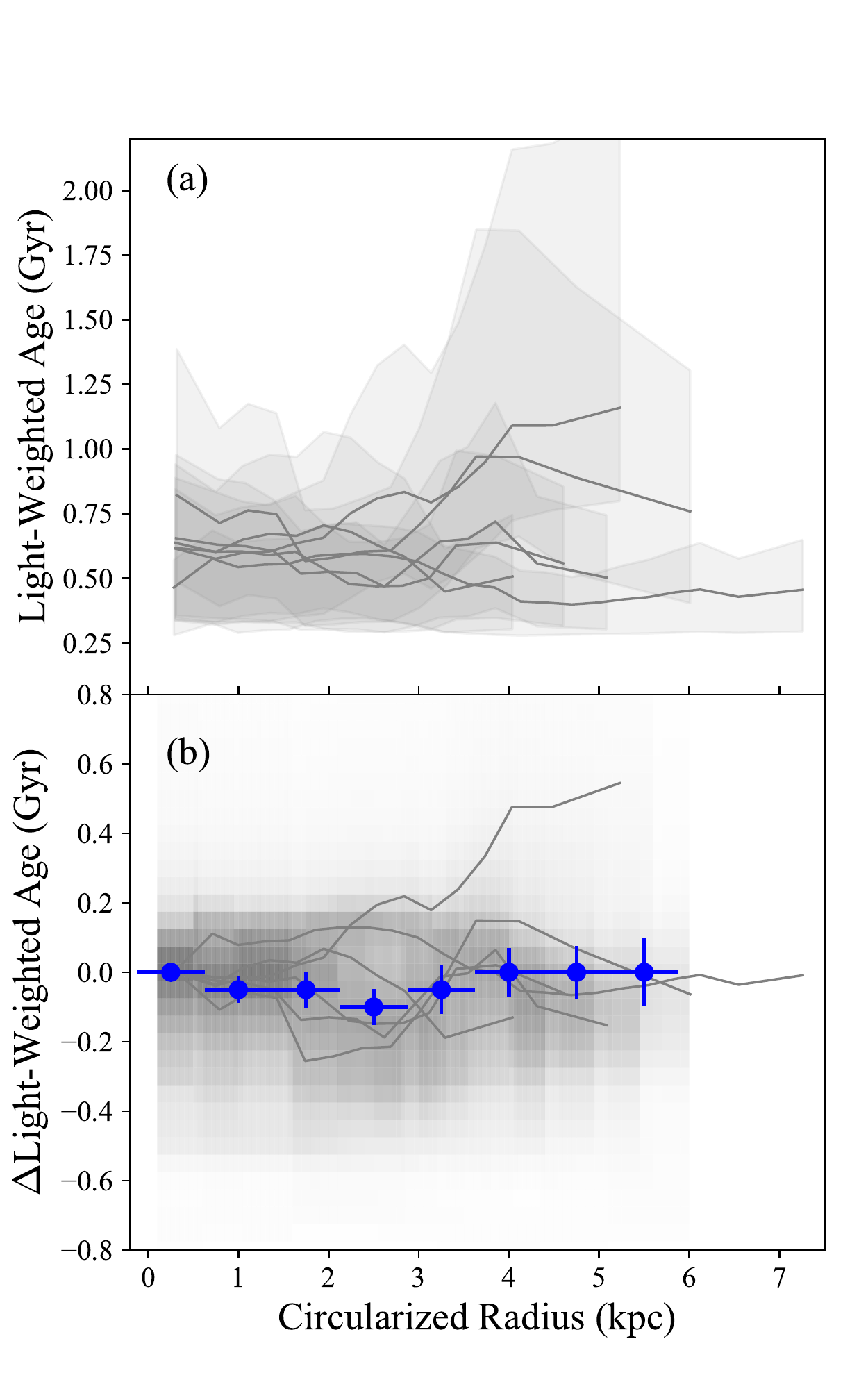}
\caption{(a) Light-weighted age at 4000$\angstrom$ as a function of the circularized radius, as derived from two-burst star formation histories. The solid lines show the median light-weighted age fit to each measurement and the grey shaded regions bound the 1-$\sigma$ spread about the median. The galaxies are consistent with being $\sim600$ Myr old. (b) Age gradient profiles as a function of the circularized radius. The shaded background represents the sum of the posteriors for the light-weighted age of all 6 galaxies in the sample, divided by the total number of galaxies in the sample. The blue points are the median of the average posterior, along with errors determined via jackknife resampling. The sample exhibits flat age gradients out to 5.5 kpc.
\label{fig:tlw}}
\end{figure}

\subsection{Flat Age Gradients in \squiggle Post-Starbursts} \label{subsec:age_gradients}

In most stellar populations, $H\delta_A$ is insufficient to constrain age because it does not monotonically increase or decrease with time and therefore cannot be inverted \citep[e.g.][]{Kauffmann2003a}. However, for very high values, $H\delta_A$ has significant constraining power because the light-weighted spectrum of a galaxy must be dominated by short-lived A-type stars to result in such extreme absorption. In this section, we utilize a simple two-burst star formation history model to constrain the radial age profiles in these galaxies. We implement this modeling in two ways. First, we treat the annular $H\delta_A$ measurements as independent, and use the model to understand the range of light-weighted ages ($t_{LW}$) that correspond to these measurements. Second, we test the extreme case of a nuclear starburst imposed on an older stellar disk to test whether the flat observed $H\delta_A$ profiles could result from an unresolved central burst. Together, these models will inform the type of intrinsic age profiles and formation mechanisms of the sample.

In order to produce model spectra, we use a simple two top hat star formation history illustrated in Figure \ref{fig:toy_model}. This model generates two distinct stellar populations, one of which is representative of a galaxy that formed stars at early times and another that represents a younger, more recently formed population. The model allows for a combination of old and young populations with the flexibility to tune the length and timing of the younger burst in addition to the mass fraction. We use the \texttt{FSPS} python package to generate composite stellar population synthesis models with custom star formation histories \citep{Conroy2009, Conroy2010,ForemanMackey2014}. We adopt \texttt{MIST} isochrones \citep{Dotter2016,Choi2016} and use \texttt{MILES} spectral libraries \citep{Sanchez-Blazquez2006, Falcon-Barroso2011}. For all models, we assume solar metallicity, and we test to ensure that our conclusions are valid for different assumptions. The model depends on 3 parameters: the time since quenching ($t_{quench}$), the secondary burst fraction ($f_{burst}$, defined as the ratio of the mass formed in the recent burst to the total mass formed), and the length of the recent burst ($t_{burst}$). We fix the older burst last for a duration of 1 Gyr centered at $z=2$ to represent star formation which occurred well before the recent burst. For an old stellar population, both $H\delta_A$ and the luminosity vary weakly with time, so our results are insensitive to the choice of the old burst's exact age and star formation history. The combination of these parameters allows for a wide range of quenching histories and naturally produces post-starburst SEDs. Example models for a burst fraction of 10\%, $t_{quench}=100$ Gyr,  and $t_{burst}=300$ Myr are shown in Figure \ref{fig:toy_model}. Because of the vastly different mass-to-light ratios, a recent burst population (blue) which only contributes a small part of the mass budget of the galaxy can still significantly dominate the light of an older population (red), resulting in a composite spectrum (green) which exhibits strong Balmer features. $H\delta_A$ is sensitive to changes in all three of the model parameters.

Using this model, we can probe the parameter space that can produce sufficiently high $H\delta_A$ to match the observations. We generate a model library with 40 linearly spaced points $0.01 \ \mathrm{Gyr} \leq t_{quench}\leq2 \ \mathrm{Gyr}$, 99 linearly spaced points with $1\% \leq f_{burst} \leq 99\%$, and 40 linearly spaced points $0.01 \ \mathrm{Gyr}\leq t_{burst}\leq \ 2 \ \mathrm{Gyr}$, and measure $H\delta_A$ and the light-weighted age (at 4000 $\angstrom$) for each star formation history. In Figure \ref{fig:binned_models}, we show the models collapsed in $f_{burst}$ versus $t_{quench}$ that can result in $H\delta_A>7 \ \angstrom$. The symbols are colored by mean light-weighted age and symbol size indicates the number of models that lie in that region of parameter space. Such high $H\delta_A$ values most often result from a high secondary burst fraction and a short time since quenching ($t_{quench}<0.8 \ \mathrm{Gyr}$), which together result in young light-weighted ages. $H\delta_A>8 \ \angstrom$ constrains the light-weighted stellar population to be younger than 1 Gyr, while $H\delta_A>9 \ \angstrom$ can only be produced by a stellar population that is between 200 and 600 Myr old. The star formation rate of the secondary burst is extremely degenerate with the length of the burst and is not well constrained by $H\delta_A$ alone. However, all models which result in $H\delta_A>7 \ \angstrom$ can at least be constrained to have star formation rates above the \citet{Whitaker2012b} star-forming main sequence at z=0.7 in order to form $10^{11}$ solar masses by the time of observation, indicating that these galaxies likely went through a recent starburst phase.

We use this library of models to fit the observed $H\delta_A$ profiles and convert empirical measurements to light-weighted age profiles and gradients. For each individual $H\delta_A$ measurement, we marginalize over $t_{quench}, f_{burst},$ and $t_{burst}$ and plot the median light-weighted age as solid lines and the 68\% confidence intervals as shaded regions in Figure \ref{fig:tlw}a. The light weighted ages in all cases are young, and in four of the six galaxies are constrained to be $\lesssim1$ Gyr at all radii. We run identical fits assuming both sub- and super- solar metallicity ($\mathrm{logZ}=[-1.0,-0.5,0.5,1.0]$) and find that systematic shifts in the inferred light-weighted ages are $\lesssim100$ Myr, which are much smaller than the errors in our fits. The shift is such that low metallicity leads to older inferred ages and high metallicity to younger ages. In Figure \ref{fig:tlw}b, we show the trends relative to the central light-weighted age. In addition, we bin the posteriors for the light-weighted age as a function of radius for the sample and show the full posterior as a shaded region in the background, with the median and errors calculated from jackknife resampling in 0.75 kpc bins as blue points and error bars. The sample average is flat, and any deviations in the median of the posterior are $\leq100$ Myr. The gradient of the average sample is flat regardless of our assumptions about metallicity.

\begin{figure*}
\includegraphics[width=\textwidth]{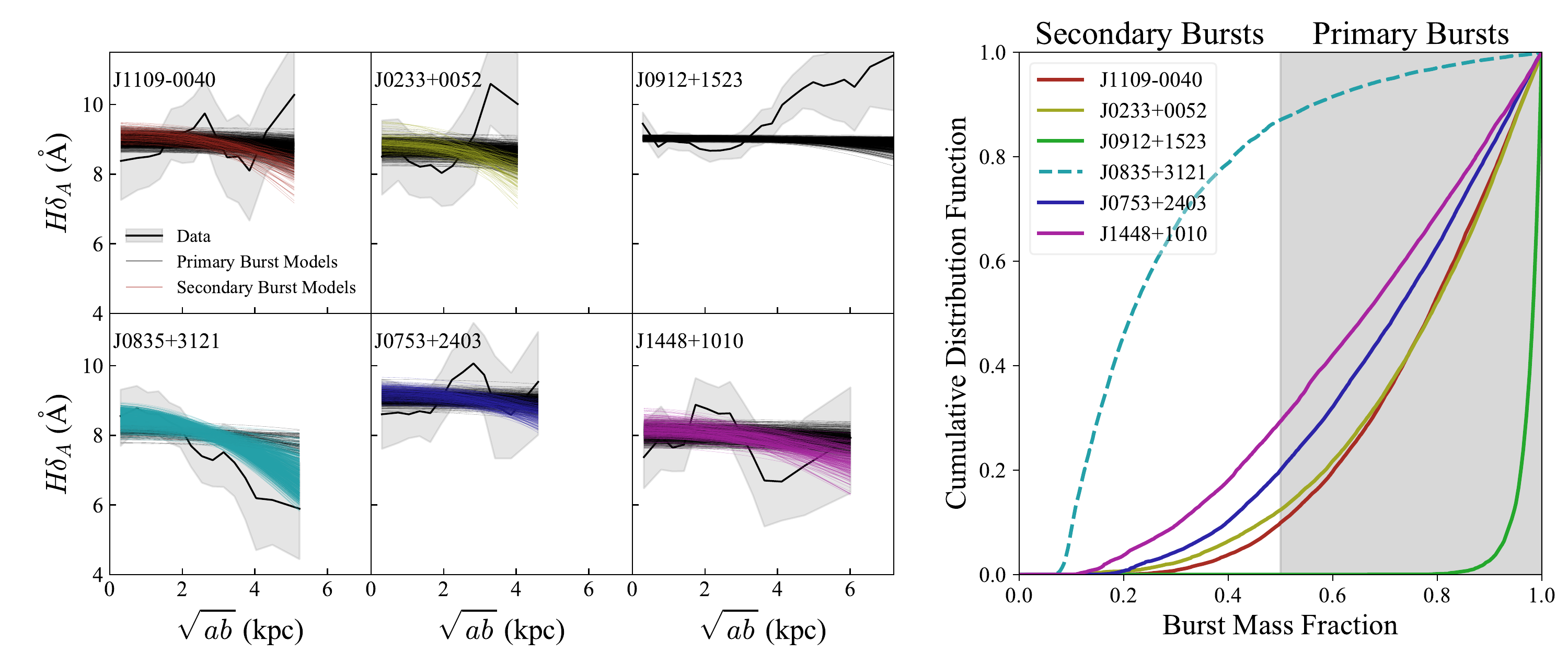}
\caption{Observed $H\delta_A$ profiles with profiles derived from central burst toy models convolved with pessimistic seeing conditions (left) and cumulative distribution function of burst fractions obtained from those models (right). 1000 models randomly drawn from the posteriors are superimposed on each panel, with colored lines indicating models that form a subdominant secondary burst and black profiles indicating models in which $>50\%$ of stars were formed in the central burst. The flat $H\delta_A$ profiles exhibited by five galaxies are only consistent with unresolved, central bursts if the majority of stars were formed in the burst. Only J0835+3121 (shown on the cumulative distribution function as a dashed line) is best fit by a secondary central burst that formed $\sim20\%$ of the total galaxy mass.
\label{fig:CDF}}
\end{figure*}

Clearly the extreme $H\delta_A$ in these massive post-starbursts necessitates that light from A-type stars dominate at all radii. However, due to the relatively low spatial resolution of our data, it is possible that this is not the result of a spatially extended post-starburst region but instead a secondary and unresolved nuclear burst of star formation. These nuclear starbursts are found in compact galaxies at $z\sim0.6$ \citep[e.g.][]{Sell2014}, and could wash out any intrinsic age-gradients by dominating the optical light under poor seeing conditions. We employ a toy model of an older, extended population superimposed with a nuclear starburst to test whether the observed $H\delta_A$ profiles could result from an unresolved secondary central burst. 

The details of this model and the fits to the galaxies are summarized in Appendix \ref{sec:centralburst}. In short, we fit the radial $H\delta_A$ profiles of each galaxy with a central young burst and a spatially-extended old population (to represent a stellar disk at $z=2$) and test whether intrinsic positive age gradients could be masquerading as flat gradients under the worst-case seeing conditions outlined in Section \ref{subsec:reduction}. The results of this fitting are shown in Figure \ref{fig:CDF} as 1000 models drawn from the posterior (left) and a cumulative distribution function for the burst fraction (right). One galaxy, J0835+3121, is well fit by a secondary central burst with $f_{burst}\sim20\%$, but the flat profiles in the remaining galaxies can only be produced by a central burst if a substantial amount of the mass (median $f_{burst}=66-98\%$, see Table \ref{tbl:centralburstfits}) was formed in the most recent episode of star formation. For these five galaxies, we conclude that the $H\delta_A$ profiles must be the result of either a spatially extended post-starburst region, or that they must have formed the majority of their mass in the last $\sim0.5$ Gyr such that the burst is not secondary, but instead is the dominant epoch of the galaxy's star formation. Either way, the optical light at all radii must be dominated by a recent burst which means that star-formation occurred and shut off uniformly throughout the galaxies.

\section{Discussion} \label{sec:discussion}

In this work, we find strong evidence that the stellar ages in $z\sim0.6$ massive post-starburst galaxies are comparable across the face of the galaxy, at least to $\sim5$ kpc. In contrast, both star-forming and quiescent galaxies tend to have intrinsically negative color gradients out to $z=2$ \citep{Suess2019}, indicating that younger stars dominate the light profiles of these galaxies at large radii while older stars dominate in their cores. In star forming galaxies, these negative age gradients can be ascribed to star formation in disks, which populates the outer regions of galaxies with younger stars \citep{Nelson2016}, and in quiescent galaxies they are likely caused by the addition of ex-situ stars, which puff up the outer regions with younger stars from less massive systems \citep{Bezanson2009, Naab2009, Hopkins2009, VanDokkum2010}. The lack of radial trends in the stellar ages of this sample indicates that if \squiggle post-starbursts are galaxies in transition from star-forming to quiescent, the process that quenches them must do so in a way that erases the existing negative age gradient from the progenitor stellar disk. The high $H\delta_A$ that we observe everywhere implies that these galaxies are not typical star-forming galaxies which simply truncated their star-formation, but instead have actually experienced an enhancement in star-formation rate which preceded quenching. Local starburst galaxies have been shown to have flat star-forming surface densities and light weighted ages \citep{Bluck2020}, and such galaxies could indeed evolve into post-starbursts like the ones we observe.

If instead the progenitors are quiescent galaxies experiencing a burst of star-formation that pushes them into the \squiggle selection, they \textit{also} must form the new stars in such a way that overcomes an age gradient that has been caused by minor mergers, and in Section \ref{sec:centralburst}, we show that this can only happen if a nuclear starburst formed $\gtrapprox50$\% of the total galaxy mass. This indicates that it is unlikely that these galaxies quenched via a halo process that cuts off the gas supply to the galaxy all at once \citep[e.g.][]{Feldmann2015}, as a uniform shutdown of star formation that does not include a significant burst would not be able to transform the existing galaxy stellar age profiles. It is also unlikely that any event that preferentially removes gas from the center of the galaxy, like AGN driven outflows \citep[e.g.][]{Mao2018}, could cause the quenching in this sample, as that would likely enhance any existing negative age gradients. Wet compaction events \citep{Tacchella2015a, Zolotov2015} are also unlikely to be responsible, as they predict negative age gradients much stronger than those we observe, which are at most consistent with $|\Delta t_{LW}|\sim100$ Myr.

One way to erase intrinsically negative age gradients is via a starburst event that is more centrally concentrated than the underlying distribution of older stars. Gas rich mergers could effectively trigger this mode of quenching by driving gas to the center of a galaxy to rapidly form a new generation of stars \citep{Hopkins2008, Snyder2011, Wellons2015}. There is significant evidence for this mode of quenching, as post-starbursts have been found to lie below the mass-size relation for both star-forming and quiescent galaxies at intermediate redshift \citep{Almaini2017,Wu2018, Wu2020}. \citet{DeugenioF2020} find evidence for positive age-gradients in the stacked $H\delta_A$ and Fe4383 profiles of the LEGA-C post-starburst sample, and \citet{Sell2014} identified a population of extreme compact starbursts in the SDSS, which have light profiles that are well-fit by an unresolved nuclear starburst superimposed with an underlying de Vaucouleurs profile. At $z=2$, star forming galaxies have been found with star forming regions which are a factor of 2 more compact than the older stellar disks, suggesting that $\sim300$ Myr depletion times would yield integrated light profiles that are similarly compact to the population of quiescent galaxies \citep{Tadaki2020}. Although one galaxy in this sample could easily be representative of this channel, in some cases the central starbursts in the \citet{Tadaki2020} sample are sufficiently extreme that the new stellar population could outshine any older stellar light, erasing age gradients. Therefore, although these star-forming galaxies are identified at an earlier epoch, we cannot rule out a low-redshift tail of the population as a possible set of progenitors of the \squiggle galaxies studied herein. 

However, in the majority of the sample, only a burst of star formation which forms the majority of the stellar mass of the galaxy or is comparably extended to the older population could result in the observed $H\delta_A$ profiles. There is evidence that photometrically-selected post-starburst galaxies at z$\sim 1-2$ exhibit flat color gradients \citep{Maltby2018, Suess2020}, which is a qualitatively similar result to the flat age gradients we detect. Furthermore, \citet{Suess2020} found that the mass-weighted sizes in post-starbursts are actually very similar to those of quiescent galaxies at a given epoch, and the difference in observed size manifests almost entirely from the accretion of stars at large radii in galaxies in post-quenching minor mergers. Since we are catching these galaxies directly after quenching, it may be that we are observing them before they have acquired their typical negative quiescent age gradient. This seems consistent with the $z\sim2$ lensed quiescent galaxies studied in \citet{Jafariyazani2020} and \citet{Akhshik2020}, which both exhibit flat age gradients and could have evolved from systems similar to the ones in this sample.

The literature includes objects that exhibit a diversity in age gradients; whether these reflect distinct quenching channels or two quenching modes that smoothly evolve in prevalence over cosmic time remains to be seen. At low-redshift, massive post-starburst galaxies appear to be quiescent galaxies that have just quenched a secondary and sub-dominant episode of star-formation. At earlier cosmic times, an older underlying population of stars does not exist, and post-starburst galaxies are galaxies that have just finished quenching their primary epoch of star formation. We posit that the \squiggle sample represents an intermediate redshift tail to the high-redshift post-starburst distribution due to their strong absorption ($\langle H\delta_A \rangle \sim7.12 \ \angstrom$) and flat age gradients, in contrast to the comparatively weaker absorption ($\langle H\delta_A \rangle \sim 5.5 \ \angstrom$) and positive age gradients in LEGA-C that more closely resemble local post-starbursts. In this paper, we have demonstrated that quenching in class of post-starburst galaxies identified in \squiggle happens simultaneously throughout the galaxies or, if centrally concentrated, is dominant in both mass \textit{and} light.

\section{Summary and Conclusions}

In this work, we study a sample of massive post-starburst galaxies at z$\sim$0.6 using spatially resolved spectroscopy. We find the following:

\begin{itemize}

    \item Three of the six galaxies show unambiguous signs of ordered motion, while the rest of the sample shows weak or unresolved ordered motion (see Figure \ref{fig:velocity_2d}). One rotating galaxy was observed under the worst seeing conditions, indicating that all galaxies are at least marginally resolved in the IFU datacubes.

    \item Five of the galaxies we observe exhibit $H\delta_A\gtrapprox7 \ \angstrom$ measured out to $r_{circ}\sim5$ kpc, indicating that an A-type stellar population dominates their optical light at all radii (see Figures \ref{fig:hdelta_2d} and \ref{fig:hdelta_all_gal}). On average, the sample exhibits flat $H\delta_A$ and light-weighted age profiles, with young ($t_{LW}\sim600 \ \mathrm{Myr}$) ages throughout (see Figure \ref{fig:tlw}).

    \item We test whether the observed $H\delta_A$ profiles could be the product of an unresolved nuclear starburst in an older quiescent galaxy. In one galaxy, we find that the observed $H\delta_A$ profile is best fit by a central burst with a secondary burst mass fraction of $\sim$20$\%$. For the remaining five galaxies, we find that their $H\delta_A$ profiles are not consistent with an unresolved central secondary starburst (see Figure \ref{fig:CDF}). 
    
    \item The finding of flat age gradients stands in contrast with other studies of less extreme post-starbursts that appear to be the products of central secondary bursts of star formation. This indicates that we have identified a sample of galaxies which have recently ended their primary epoch of star formation in a way that quenches the entire galaxy within $\sim100$ Myr.
    
\end{itemize}
    
The fundamental limitation of this study is the seeing, which, from the ground, is comparable to the sizes of the galaxies at this redshift. Future work using adaptive optics or space based IFU such as NIRSPEC on the James Webb Space Telescope could probe galaxies in transition with finer resolution. JWST in particular would have the advantage of pushing out to IR wavelengths where any residual star formation can be spatially resolved with $H\alpha$. Although spectroscopic identification of post-starburst galaxies is optimal, identification of galaxies within SDSS limits the \squiggle sample to the tail end of a post-starburst distribution that peaks at earlier times \citep{Whitaker2012b, Wild2016}. Future large surveys like the Dark Energy Spectroscopic Instrument \citep{DESI} and Prime Focus Spectrograph \citep{Takada2014} surveys will allow for spectroscopic identification of post-starburst galaxies at $z>1$, when we expect the rapid quenching process to be more dominant. Future studies with these exciting new samples and instruments will continue to improve our understand of the gas, kinematics, and stellar populations of these higher redshift post-starburst galaxies to understand how galaxies transform during the peak epoch of quenching.

    
\acknowledgements 
We thank the referee for a very helpful report
that improved this manuscript. DS, JEG, RSB, and DN gratefully acknowledge support from NSF grant AST1907723. This work was performed in part at Aspen Center for Physics, which is supported by National Science Foundation grant PHY-1607611. DS would also like to thank the North American ALMA Science center for financial support in attending the ACP ``Quenching" conference. RF acknowledges financial support from the Swiss National Science Foundation (grant no 157591 and 194814). DS would also like to thank Alan Pearl for all his program wisdom help along the way, along with the entire Pitt Galaxies Group for looking at infinite iterations of figures. This research made use of \texttt{pyphot} to measure Lick Indices on spectra, an open-source package for Python hosted at https://github.com/mfouesneau/pyphot

This work was based on observations obtained at the international Gemini Observatory, a program of NSF’s NOIRLab, which is managed by the Association of Universities for Research in Astronomy (AURA) under a cooperative agreement with the National Science Foundation. on behalf of the Gemini Observatory partnership: the National Science Foundation (United States), National Research Council (Canada), Agencia Nacional de Investigaci\'{o}n y Desarrollo (Chile), Ministerio de Ciencia, Tecnolog\'{i}a e Innovaci\'{o}n (Argentina), Minist\'{e}rio da Ci\^{e}ncia, Tecnologia, Inova\c{c}\~{o}es e Comunica\c{c}\~{o}es (Brazil), and Korea Astronomy and Space Science Institute (Republic of Korea). The data was obtained under the following observing programs: GN-2016A-FT-6, GN-2017B-Q-37, GS-2018A-FT-112, and GN-2019A-Q-234. DS would also like to thank the Gemini Helpdesk for their assistance in the reduction of the GMOS data.

Funding for SDSS-III has been provided by the Alfred P. Sloan Foundation, the Participating Institutions, the National Science Foundation, and the U.S. Department of Energy Office of Science. The SDSS-III web site is http://www.sdss3.org/. 

SDSS-III is managed by the Astrophysical Research Consortium for the Participating Institutions of the SDSS-III Collaboration including the University of Arizona, the Brazilian Participation Group, Brookhaven National Laboratory, Carnegie Mellon University, University of Florida, the French Participation Group, the German Participation Group, Harvard University, the Instituto de Astrofisica de Canarias, the Michigan State/Notre Dame/JINA Participation Group, Johns Hopkins University, Lawrence Berkeley National Laboratory, Max Planck Institute for Astrophysics, Max Planck Institute for Extraterrestrial Physics, New Mexico State University, New York University, Ohio State University, Pennsylvania State University, University of Portsmouth, Princeton University, the Spanish Participation Group, University of Tokyo, University of Utah, Vanderbilt University, University of Virginia, University of Washington, and Yale University.

\appendix

\begin{deluxetable*}{cccccccccccc}
\tabletypesize{\scriptsize}
\tablecaption{\label{tbl:centralburstfits}}
\tablehead{
    \colhead{Name} & \colhead{Max FWHM\tablenotemark{1}} & \multicolumn{3}{c}{$t_{quench}$ (Gyr)} & \multicolumn{4}{c}{Burst Mass Fraction} & \multicolumn{3}{c}{$t_{burst}$ (Gyr)} \\[-0.2cm]
    \colhead{} & \colhead{["]} & \colhead{16$\%$} & \colhead{50$\%$} & \colhead{84$\%$} & \colhead{16$\%$} & \colhead{50$\%$} & \colhead{84$\%$} & \colhead{5$\%$} & \colhead{16$\%$} & \colhead{50$\%$} & \colhead{84$\%$}
}

\startdata
J1109-0040 & 0.64 & 0.06 & 0.16 & 0.41 & 0.57 & 0.79 & 0.94 & 0.43 & 0.35 & 0.76 & 1.03 \\
J0233+0052 & 0.57 & 0.05 & 0.15 & 0.42 & 0.54 & 0.79 & 0.94 & 0.37 & 0.3 & 0.84 & 1.17 \\
J0912+1523 & 0.51 & 0.06 & 0.18 & 0.41 & 0.96 & 0.98 & 1.0 & 0.92 & 0.42 & 0.83 & 1.05 \\
J0835+3121 & 0.82 & 0.09 & 0.22 & 0.42 & 0.12 & 0.22 & 0.46 & --\tablenotemark{2} & 0.2 & 0.57 & 1.1 \\
J0753+2403 & 0.79 & 0.07 & 0.18 & 0.41 & 0.46 & 0.72 & 0.92 & 0.32 & 0.27 & 0.68 & 0.95 \\
J1448+1010 & 0.89 & 0.05 & 0.18 & 0.43 & 0.38 & 0.66 & 0.9 & 0.22 & 0.42 & 1.13 & 1.66 
\enddata

\tablenotetext{1}{See Section \ref{subsec:reduction} for details on the upper limits on the full-width half maximum of the point spread function}
\tablenotetext{2}{For J0835+3121, we do not quote the 5\% value in the burst fraction because the distribution is not single tailed from a burst mass fraction of 1}
\end{deluxetable*}

\section{Unresolved Central Burst Toy Model} \label{sec:centralburst}

In order to test whether the observed flat $H\delta_A$ could be the result of an unresolved nuclear starburst, we assume a simple geometry for an underlying stellar population that formed at $z\sim2$ and superimpose a central burst under the worst case seeing conditions. We distribute the light from the older population following an exponential disk with $r_e$=3 kpc, the characteristic size of a late type $\sim 10^{10.5} M_{\odot}$ galaxy at $z\sim2$. To this underlying profile we add a pointlike central starburst, convolved with a Moffat profile with the conservative FWHM limit (see Section \ref{subsec:reduction}).

\begin{figure*}
\includegraphics[width=\textwidth]{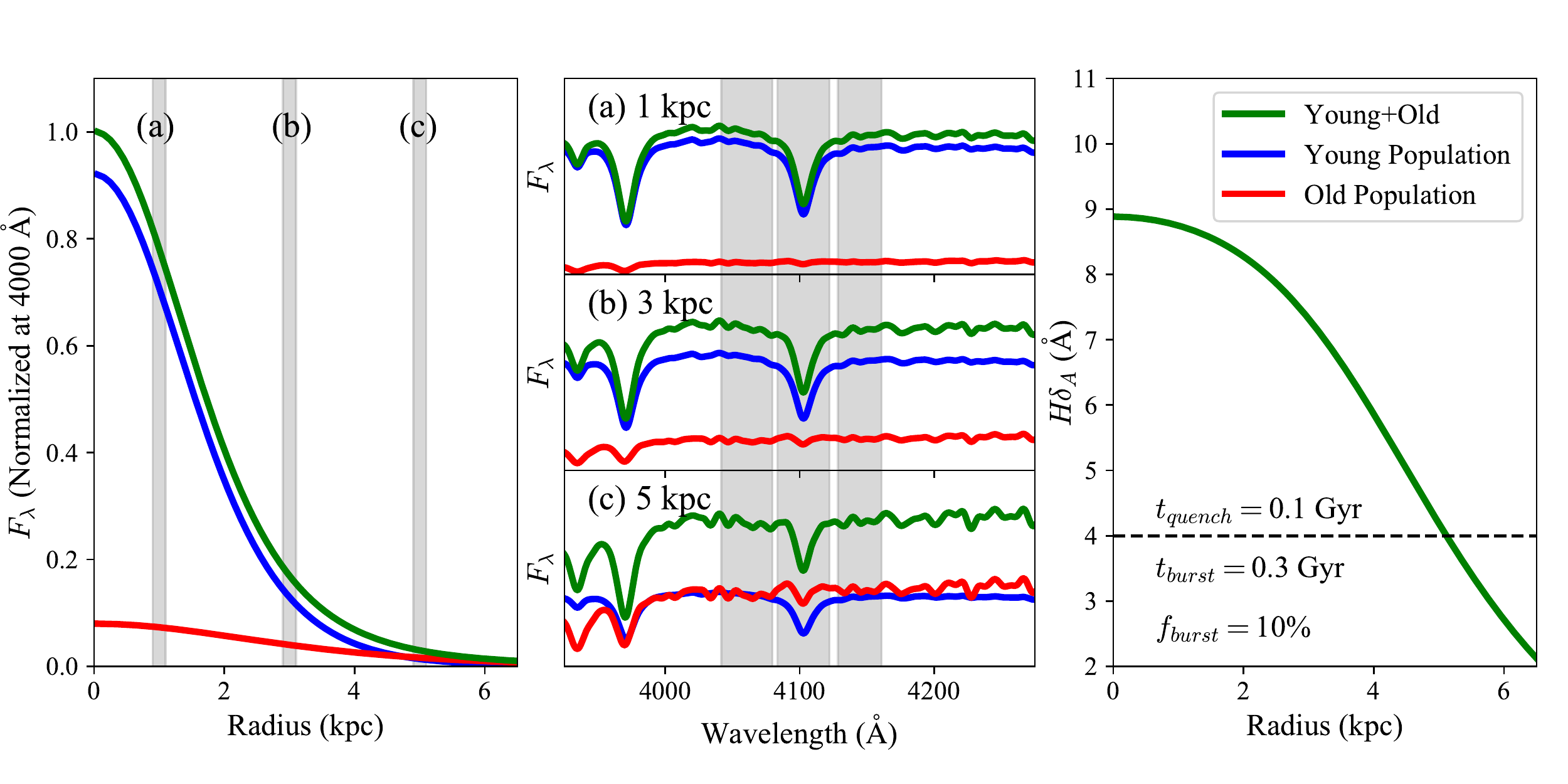}
\caption{An example of the radial variation in $H\delta_A$ we would see for an unresolved central starburst and an underlying older population in an $r_e$=3 kpc exponential disk profile, using the same parameters as Figure \ref{fig:toy_model}. (Left): The normalized intensity profiles for the populations. The old profile is convolved with the Moffat seeing and is plotted in red. The burst profile is a Moffat profile with FWHM of 0.5" and is plotted in blue, and the green profile is the sum of the two. The grey bands labeled a, b, and c are located at 1, 3, and 5 kpc respectively. (Center): The evolution of the SEDs in the $H\delta_A$ bandpass at 1, 3, and 5 kpc. The coloring convention is the same as the previous figure, where red is the older spectrum, blue is the recent starburst spectrum, and green is the composite spectrum. The spectra are normalized to the flux at 4000 $\angstrom$ in the composite spectra. At low radius, the central burst dominates the light, but at 5 kpc, the older population is contributing non-negligible flux, resulting in weaker $H\delta_A$ absorption. The gray bands indicate the two continua for the $H\delta_A$ Lick index, as well as the line range. (Right): The $H\delta_A$ profile that would result from this model, which falls off at large radii.
\label{fig:model_demo}}
\end{figure*}

The resulting intensity and $H\delta_A$ profiles for the same burst as in Figure \ref{fig:toy_model} are shown in Figure \ref{fig:model_demo} for a galaxy observed under 0.5" seeing conditions. The left panel shows the intensity profile in the B band, which is dominated by an A-type stellar population well past the half-width-half-maximum of the Moffat central burst due to the much smaller mass-to-light ratio of the recently formed stars. The coloring scheme is the same as in Figure \ref{fig:toy_model}, where the blue light profile corresponds to the young population, red to the old population, and green to the composite. The labeled bands in the left panel correspond to the 3 sub-panels in the center of the figure, which show the spectral region around $H\delta$ at 1, 3, and 5 kpc. In the outer radii, light from the older population becomes more dominant, and by 5 kpc the older population is contributing more light to the $H\delta_A$ feature than the recent secondary burst. The $H\delta_A$ profile for this model configuration is shown in the right panel and by $\sim$ 5 kpc the absorption has fallen below the common post-starburst selection of $H\delta_A >4 \ \angstrom$.

For each galaxy in this sample, we fit the 3-parameter central burst model to the $H\delta_A$ profiles using emcee \citep{Foreman-Mackey2013}. We assume the following flat priors on our parameters: $0.001 \ \mathrm{Gyr} <t_{quench}<2 \ \mathrm{Gyr}$ and 0$<f_{burst}<$1. We allow the length of the burst, $t_{burst}$, to be anywhere from 0.001 Gyr to the maximum time between quenching and the end of the old burst. We fit the models to the $H\delta_A$ profile and integrated measurements for each galaxy. We run the fits using 24 walkers and 5000 iterations and exclude the first 500 iterations to ensure burn in. Visual inspection of our walkers confirms that the fits have converged. The resulting best fitting parameters are shown in Table \ref{tbl:centralburstfits}.

\begin{figure*}
\plotone{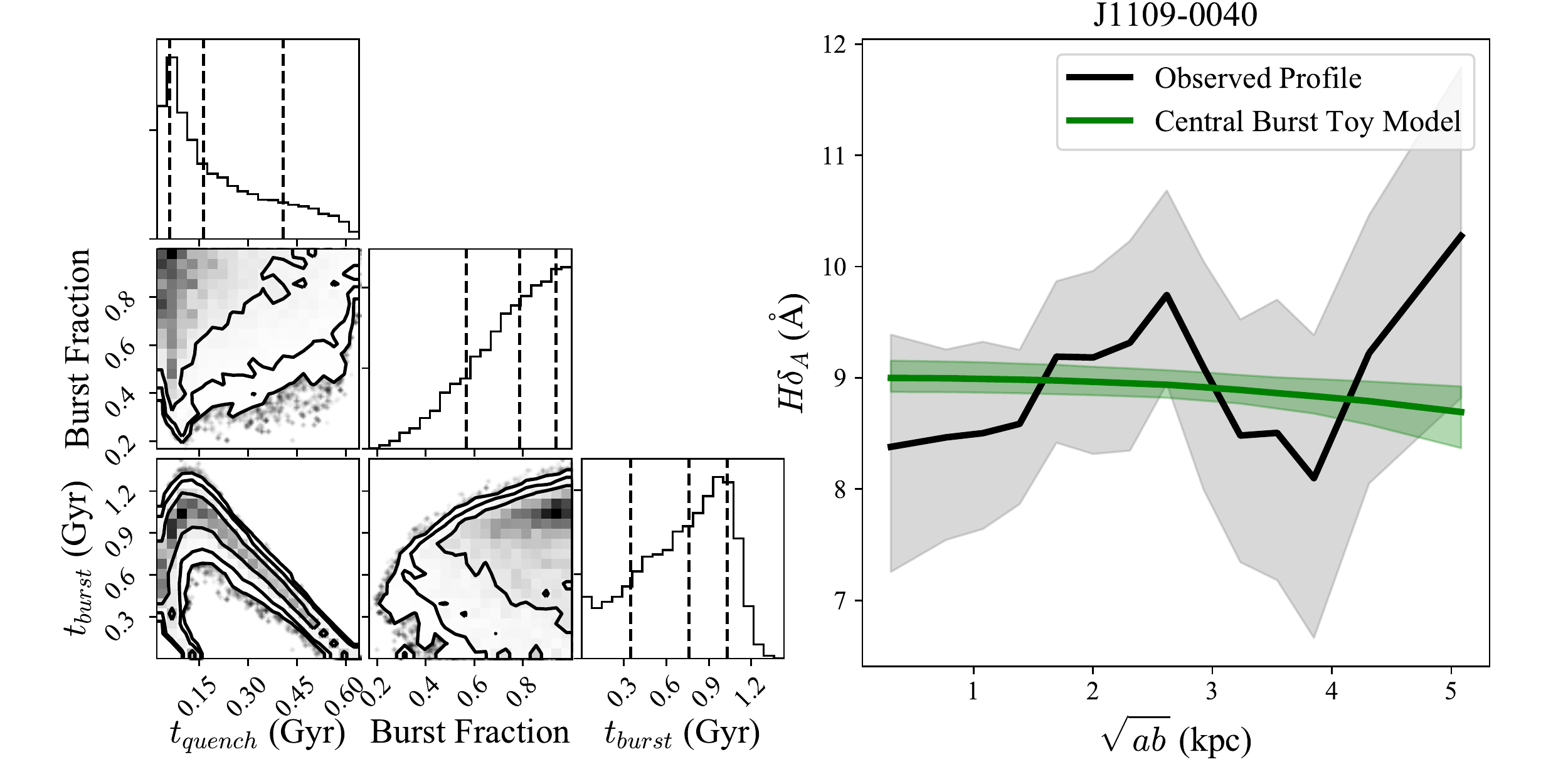}
\caption{The results of fitting our two-burst model to J1109-0040. (Left): The corner plot resulting from our MCMC run showing the posterior of our 3 parameter model. The contours represent 1, 2, and 3 sigma confidence intervals, and the dashed lines on the 1D histograms are the median and upper and lower 68\% regions. Note that the burst fraction 2-D histogram is physically bounded between 0 and 1, so it is not concerning that our fit to a flat profile runs up against that boundary. (Right): The $H\delta_A$ profile for this galaxy with the median and upper and lower 68\% models plotted on top, showing that we are able to achieve a good fit to the observed profile. This modeling allows us to place strong constraints on the central burst strength below which we would expect our profile to be distinct from the flat profile we observe.
\label{fig:0278_model}}
\end{figure*}

In Figure \ref{fig:0278_model}, we show the results of the fitting on one of the galaxies with a flat $H\delta_A$ profile, J1109-0040. The left panel shows the corner plot for the 3 model parameters with 68-, 95-, and 99-percent contours bounding the data. The right panel shows the observed $H\delta_A$ profile along with the median and 1-$\sigma$ bounds in the best fitting model generated from 1000 random draws from the posterior distributions. A central burst can still be hidden by poor seeing conditions, but only if the majority of the mass of the galaxy formed in the later burst. This essentially rules out a central \textit{secondary} burst of star formation; the only solutions that fit the data at the 1-$\sigma$ level require $f_{burst}\geq 50\%$.

\begin{figure*}[]
\plotone{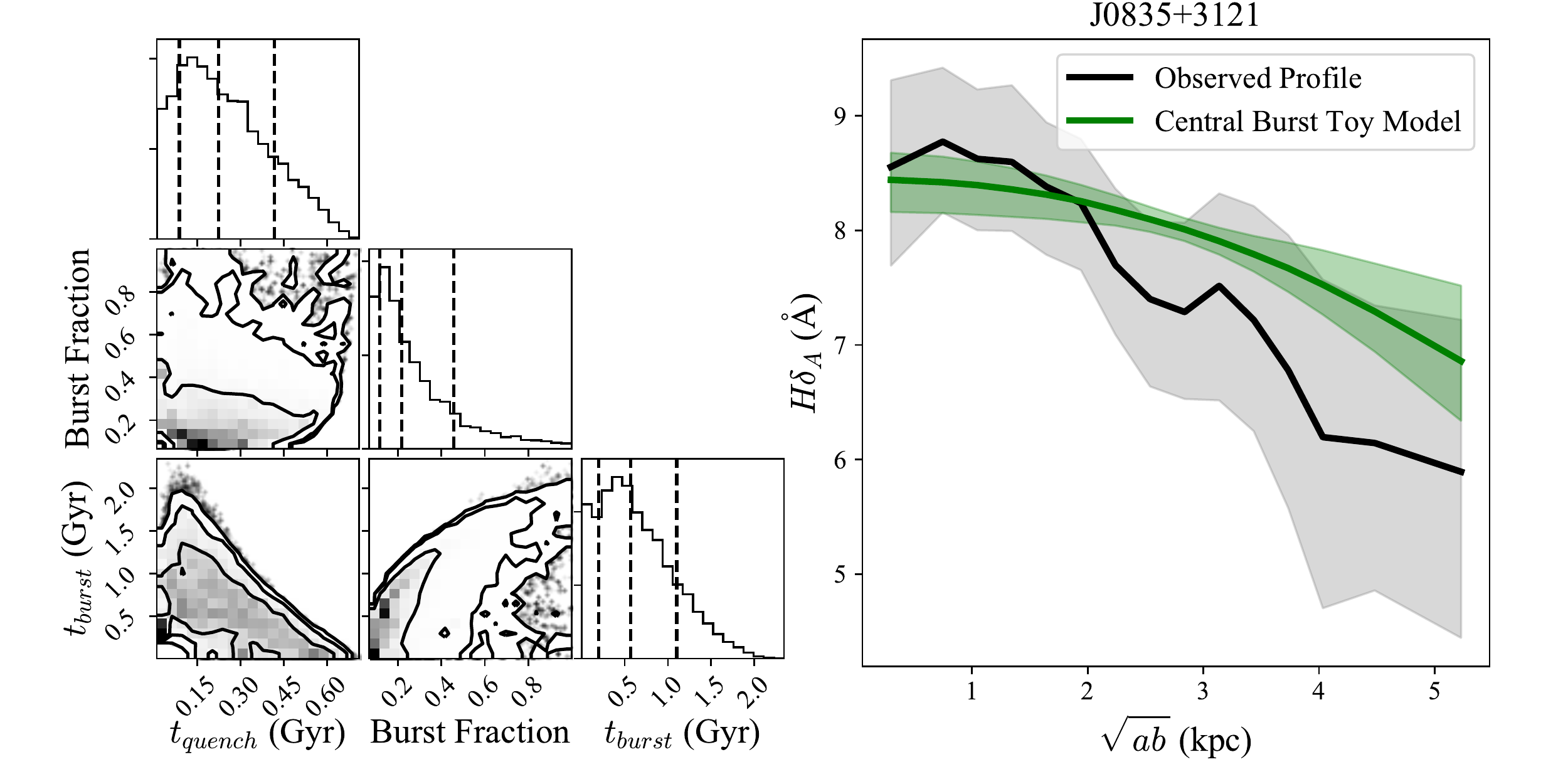}
\caption{The results of fitting our two-burst model to J0805+3121. (Left): The corner plot resulting from our MCMC run showing the posterior of our 3 parameter model. The contours represent 1, 2, and 3 sigma confidence intervals, and the dashed lines on the 1D histograms are the median and upper and lower 68\% regions. (Right): The $H\delta_A$ profile for this galaxy with the median and upper and lower 68\% models plotted on top, showing that we are able to achieve a good fit to the observed profile. In contrast to J1109-0040, here the fits allow us to place an upper limit on the central burst fraction that would agree with the observed declining $H\delta_A$ profile. 
\label{fig:4449_model}}
\end{figure*}

In Figure \ref{fig:4449_model}, we show the results of the fitting for for J0835+3121, which prefers a lower secondary burst fraction in contrast to the other 5 galaxies in our sample. For this galaxy, we place a rough upper limit on the fraction of stars that formed in a central burst: above this fraction, we would expect our observed profile to be significantly more flat than the observed profile. It appears that our worst case PSF may over-estimate the true PSF, which is why the models are not able to decline as quickly as the observed profile. Interestingly, we detect ordered motion in this galaxy (see Figure \ref{fig:velocity_2d}), and while we cannot rule out that a merger is the cause of that motion, it is clear that the galaxy is not entirely dispersion-dominated at this stage of its evolution.

\begin{figure*}[htb]
\begin{center}
{
\includegraphics[width=.48\linewidth]{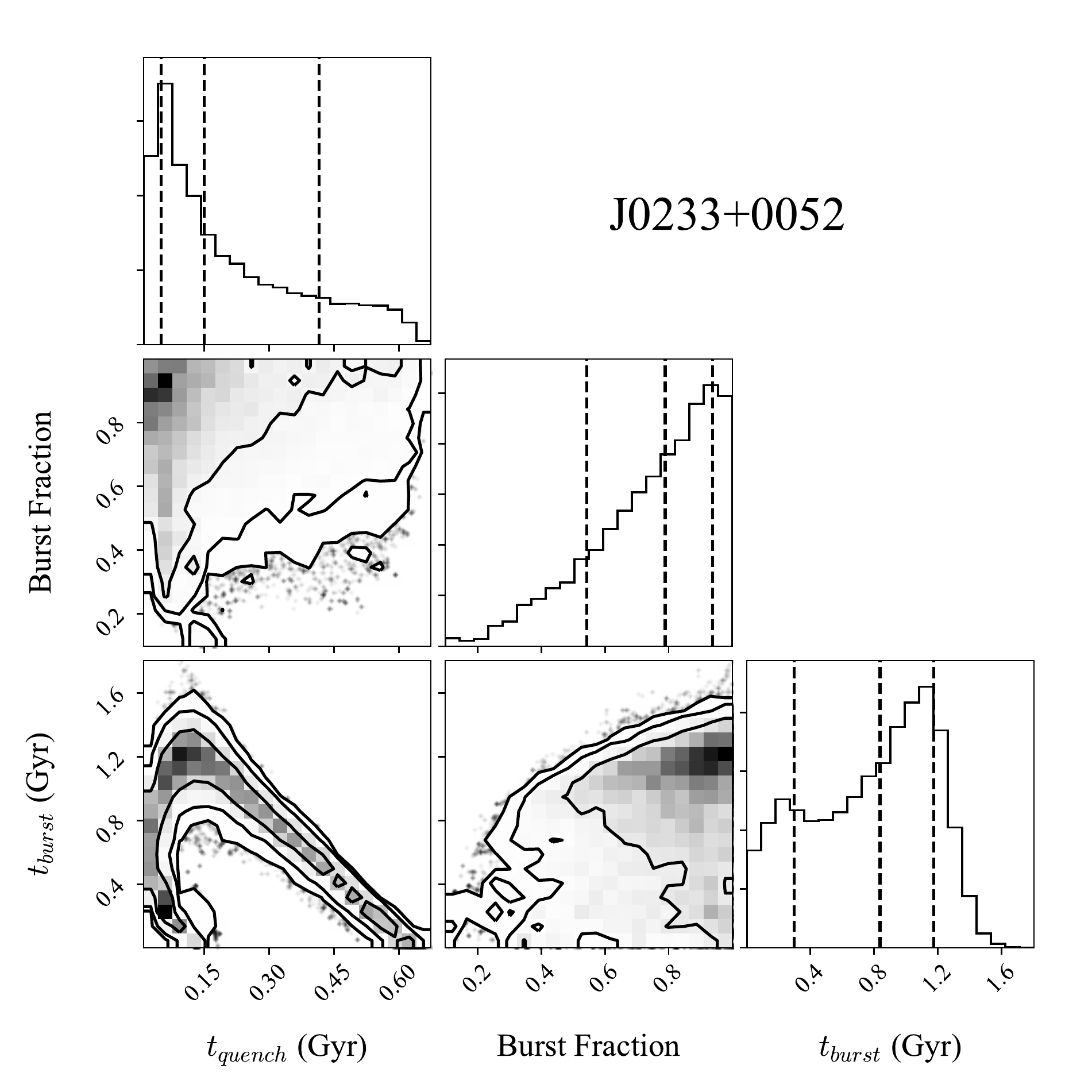}
}\quad
{
\includegraphics[width=.48\linewidth]{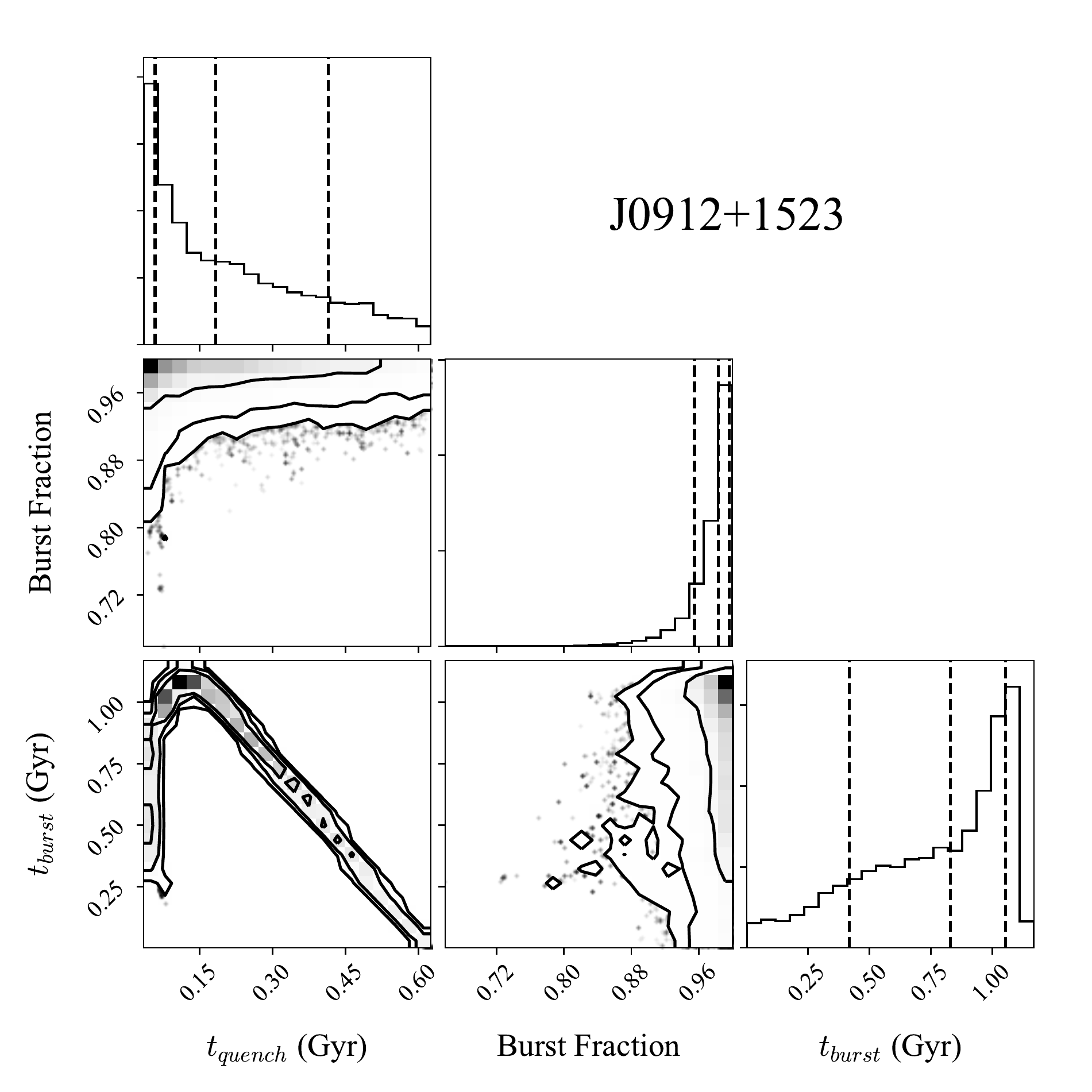}
}\\
{
\includegraphics[width=.48\linewidth]{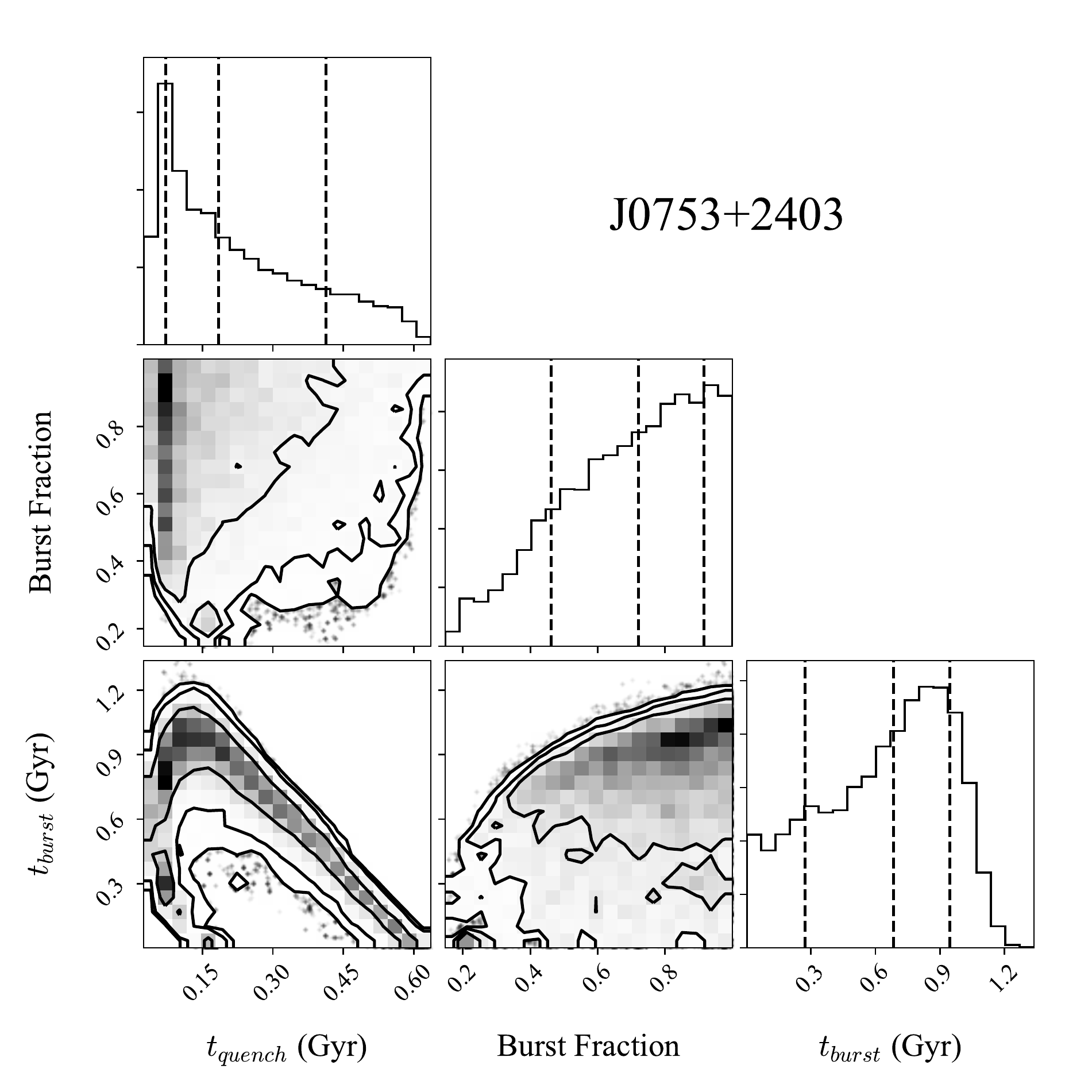}
}\quad
{
\includegraphics[width=.48\linewidth]{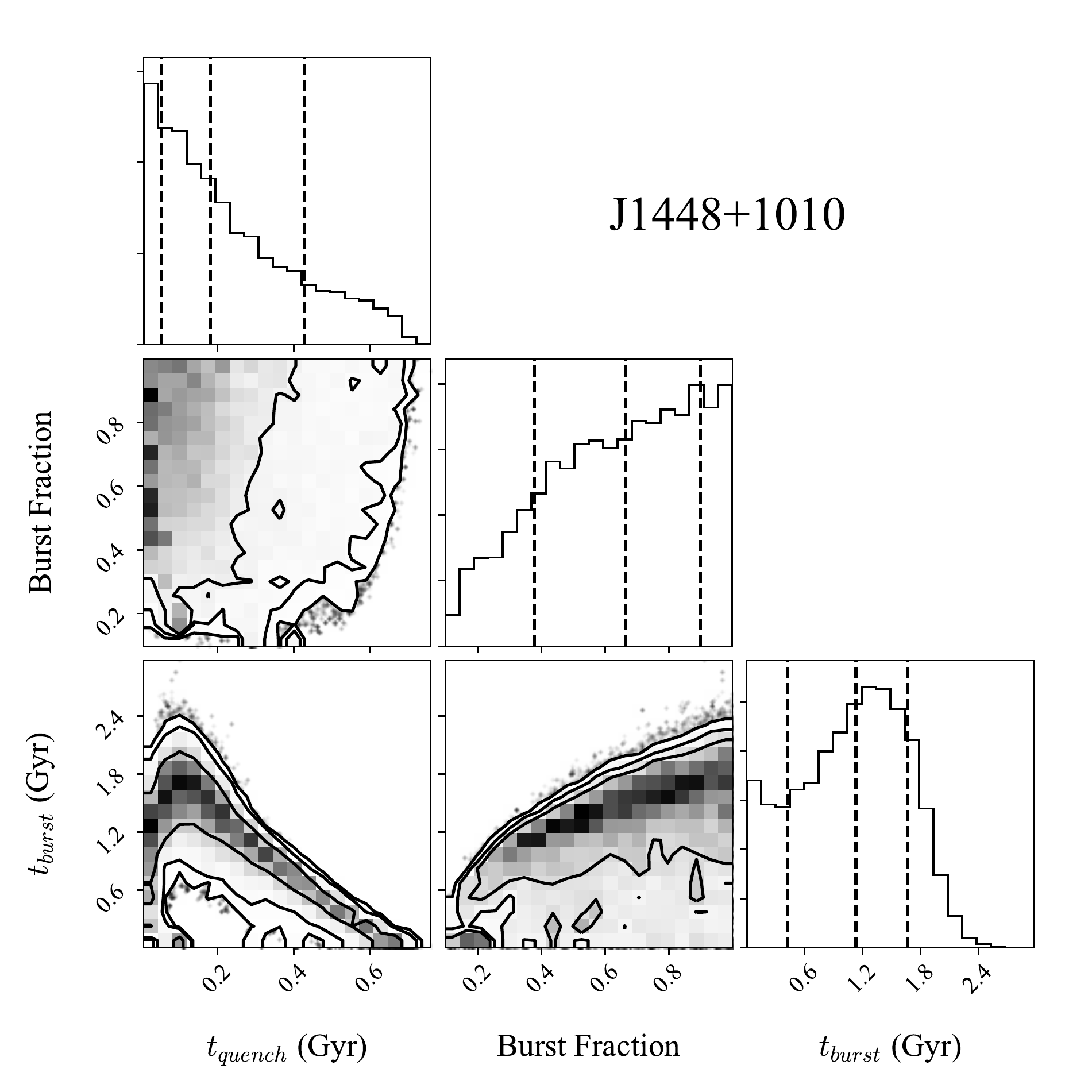}
}
\caption{The posteriors for the four galaxies not shown in Figures \ref{fig:0278_model} and \ref{fig:4449_model}. The unresolved central starburst model can only match the profiles of these galaxies with very high burst fractions.\label{fig:all_corner}}
\end{center}
\end{figure*}

The posteriors for the four remaining galaxies are shown in Figure \ref{fig:all_corner}. All four of them are similar to J1109-0040 in that the burst fraction can be constrained as a single tailed distribution from a 100\% burst, which will result in a perfectly flat profile. The exact constraints vary between the galaxies, with extreme cases like J0912+1523 being entirely inconsistent with an unresolved secondary starburst. Other galaxies are less constrained due to less extended profiles or worse seeing conditions, but they still can only produce the observed profiles if the central burst is comparable in mass to any underlying population.

\end{document}